\documentclass[sigconf,nonacm]{acmart}

\usepackage{multirow}
\usepackage{booktabs}
\usepackage{color, colortbl}
\usepackage{placeins}
\usepackage{graphicx}
\usepackage{hyperref}
\usepackage{balance}
\usepackage{xurl}
\usepackage{fontawesome}
\usepackage[position=top]{subfig}
\usepackage{tikz}
\usepackage{threeparttable}
\usepackage{nicematrix}
\usepackage{enumitem}
\usepackage{pdfpages}
\newlist{multiselect}{itemize}{2}
\setlist[multiselect]{label=$\square$}
\usepackage{multicol}

\begin{document}

\title[Influences of Displaying Permission-related Information on Web Single Sign-On Login Decisions]{Influences of Displaying Permission-related Information\\on Web Single Sign-On Login Decisions}
\titlenote{Journal version is to appear in \textit{Computers \& Security}. DOI: \url{https://doi.org/10.1016/j.cose.2023.103666}. \textcopyright 2023. This manuscript version is made available under the CC-BY-NC-ND 4.0 license \url{https://creativecommons.org/licenses/by-nc-nd/4.0/}}

\begin{abstract}
Web users are increasingly presented with multiple login options, including password-based login and common web single sign-on (SSO) login options such as ``Login with Google'' and ``Login with Facebook''.
There has been little focus in previous studies on how users choose from a list of login options and how to better inform users about privacy issues in web SSO systems.
In this paper, we conducted a 200-participant study to understand factors that influence participants' login decisions, and how they are affected by displaying permission differences across login options; permissions in SSO result in release of user personal information to third-party web sites through SSO identity providers.
We compare and report on login decisions made by participants before and after viewing permission-related information, examine self-reported responses for reasons related to their login decisions, and report on the factors that motivated their choices.
We find that usability preferences and inertia (habituation) were among the dominant factors influencing login decisions.
After participants viewed permission-related information, many prioritised privacy over other factors, changing their login decisions to more privacy-friendly alternatives.
Displaying permission-related information also influenced some participants to make tradeoffs between privacy and usability preferences.
\end{abstract}


\author{Srivathsan G. Morkonda}
\affiliation{
  \institution{Carleton University}
  \city{Ottawa}
  \country{Canada}}
\email{srivathsan.morkonda@carleton.ca}

\author{Sonia Chiasson}
\affiliation{
  \institution{Carleton University}
  \city{Ottawa}
  \country{Canada}}
\email{chiasson@scs.carleton.ca}

\author{Paul C. van Oorschot}
\affiliation{
  \institution{Carleton University}
  \city{Ottawa}
  \country{Canada}}
\email{paulv@scs.carleton.ca}

\settopmatter{printfolios=true}
\maketitle

\section{Introduction}
Many proposed password-replacement schemes aim to overcome the security and usability issues related to user-chosen passwords~\cite{chiasson2009multiple,ur2015added}. These in turn typically have their own limitations related to usability and deployability that hinder their adoption~\cite{bonneau2012quest}.
As a promising alternative, web Single Sign-On (SSO) schemes offer strong security benefits while also addressing many of the challenges prevalent in other proposals~\cite{alaca2020comparative}.

Web SSO schemes are increasingly common in consumer web services alongside traditional password-based login methods.
These SSO schemes aim to improve login security and reduce the number of passwords users need to remember as they enable authentication to \textit{relying party} (RP) sites using a single \textit{identity provider} (IdP) account.
Besides these user benefits, common SSO schemes such as OAuth 2.0~\cite{oauth2rfc} are desirable to service providers as they enable RP sites to request and obtain access to user personal information through APIs offered by IdP platforms.
In exchange, these RPs could offer users additional functionality and usability improvements using the personal information obtained from IdP APIs, though the extra data could also be used for marketing and tracking purposes.

Many RP sites that support SSO login offer multiple login options,  often with popular IdPs such as Google, Facebook, and Apple~\cite{morkonda2021empirical,jarpehult2022longitudinal}.
These sites prompt users to choose from a list of multiple SSO and non-SSO (e.g., password-based) login options where each login choice might involve different user benefits and costs.
For example, an RP login prompt with non-SSO login (e.g., with username and password) and SSO login (e.g., with Google) options might request access to a user's list of contacts (if SSO login is chosen) to allow the user to share content to a contact via email without the user being required to manually enter the target email address, thereby favouring either convenience (of SSO) or privacy (of non-SSO).
RP login prompts usually contain little or no information about the available choices, with respect to the privacy and security implications of choosing a given login option.
Given the lack of transparency at the time when users are making a choice among available options, users might inadvertently choose an option that is misaligned with their personal preferences (or interests) as a result of being not fully informed about the implications of using SSO login.

Previous user studies~\cite{balash2022security,robinson2014cognitive,bauer2013comparison,egelman2013my} involving SSO usability, privacy, and security focus on mental models and user preferences for individual IdP platforms, providing insights into user behaviour after they have made the login choice to use a given IdP login option.
In this paper, we focus on the research gap of understanding what influences users when they must choose from a list of login options presented on RP sites, and on the effect of providing privacy information in advance of, users' login choices.
To this end, we pursue the following research questions:
\begin{itemize}
    \item[RQ1:] What factors influence users' login decisions when presented with a list of SSO and non-SSO login options?
    \item[RQ2:] What effect (if any) does displaying comparative privacy information about SSO choices have on users' login decisions?
\end{itemize}

\noindent
Our goal in RQ2 is to test whether users would make more privacy-aware choices (e.g., switch to a privacy-friendly login) when given information about the user data requested by RPs through individual SSO login options.
In a related paper~\cite{morkonda2022ssoprivateeye}, we describe a browser extension tool we built to inform user login decisions; our tool augments the current SSO workflow---without requiring cooperation from RPs---to display to users comparative information about the SSO user data requests. Sec.~\ref{sec.methodology.ssoprivateeye} gives further details on the tool.
In the present paper, we investigate users' login choices in the current SSO workflow, and compare these choices to how users would choose in an augmented workflow, i.e., if the comparative information on RP permissions was available at the start of the login workflow.

We conducted a user study wherein 200 participants were asked to choose their preferred login options among a list of different SSO and non-SSO login choices on a given RP site.
We then displayed to participants privacy-related information about the permissions requested by RPs across individual IdP login options, and compared the login decisions before and after this information was presented.

In examining participant choices made in the current workflow (where nothing more than a list of SSO and non-SSO login options is presented), we found that among SSO users, login with Google was significantly more popular than Facebook or Apple logins. 
Usability preferences and inertia (based on existing habits) were among the popular factors influencing these decisions.
Participants who prioritised usability had a tendency to choose SSO login options, while conversely, those prioritising security or privacy preferred non-SSO login options.
After presenting the augmented workflow, where privacy-related information is given before prompting for a login choice, participants' login decisions shifted towards more privacy-friendly login options as they prioritised privacy over other factors.
As another effect of displaying this information, some participants indicated that they consciously made a tradeoff between privacy benefit (or cost) and usability cost (or benefit).

We believe that our findings can help inform web SSO stakeholders, including RPs, IdPs, and browser vendors, to build login systems with permissions that are more transparent to users, and guide privacy practitioners to design privacy-enhancing interfaces and browser extensions or other tools that inform users about relevant privacy issues in web SSO login systems.

\section{Background on SSO and IdP design}
Here we provide background on web SSO workflow, and discuss the security and privacy design of major IdP (Google, Facebook, and Apple) platforms.

\subsection{Web Single Sign-On}
\label{sec.background.websso}
The widespread adoption of SSO has been driven in part by the ease with which users can login to RP sites using SSO.
On a typical RP site that uses OAuth-based SSO protocol (e.g., OAuth 2.0~\cite{oauth2rfc}, OpenID Connect 1.0~\cite{openIdConnect1Spec}), when a user clicks on an SSO login option (such as ``Login with Google''), the RP sends an OAuth \textit{authorization request} by redirecting the user's browser agent to an IdP login page (Google).
As part of this request, the RP specifies a list of user data attributes (in the OAuth \texttt{scope} parameter~\cite{oauth2rfc}) that it wants to access about the user from the IdP.
The IdP processes the request by first asking the user to authenticate by entering their IdP account credentials (if they are not already logged in with the IdP), and then showing the list of user data attributes requested by the RP and prompting the user to approve the RP's request.

If the user approves one or more permissions, the IdP redirects the user back to the RP (at a pre-registered landing page) along with an OAuth \textit{access token} or \textit{authorization code} (depending on the OAuth \textit{grant type} in use) that grants the RP access to user data approved by the user.
At this point, the user is said to be logged in with the RP site which can retrieve user data from the IdP and use it to provide RP services such as a more personalised user experience (e.g., show local services based on the user's current location).

\subsubsection{\textbf{Multiple SSO login options}}
\label{sec.background.multiplesso}
Many RP sites offer multiple login choices with different IdPs (e.g., Google, Facebook, or Apple).
However, they might request different amounts of personal information about the user from the individual IdPs, often with one option more privacy invasive than others~\cite{morkonda2021empirical}.
This means that SSO users on a given RP site might view (and grant) more permissions than necessary to use the RP site by selecting an SSO login option that requests more permissions compared to alternate login choices offered by the RP.
These privacy differences across the different login options remain hidden, unless the user chooses to login with each SSO option and compare the requested permissions.

Manually collecting and comparing these permissions is time-consuming as there are significant usability costs involved in completing login with each listed SSO login option before viewing the IdP consent dialog.
This lack of visibility into SSO permissions in current UI designs means that users might make login decisions and subsequently grant SSO permissions without full information.

Description of permissions in IdP consent dialogs tend to be vague and insufficient to adequately inform users about the privacy implications of using SSO to login~\cite{robinson2014cognitive, balash2022security}.
The necessity of permissions might not be visible as users on an RP site with different SSO login options are only shown the permissions requested with the chosen login option.
This design could lead to SSO users of certain RP sites incorrectly assuming that the requested permissions are mandatory for the RP site to provide its services, which may not be clear until they are fully informed about alternate privacy choices.

These issues motivate us to study the factors that influence login decisions and the effects of presenting comparative information about the IdP permissions.

\subsection{Overview of IdP security and privacy design}
We now discuss three popular IdP platforms (Google, Facebook, and Apple) and their designs related to SSO privacy and usability.

\subsubsection{\textbf{IdP platform differences}}
IdP platforms including Facebook, Google, and Apple offer different SSO features that could impact privacy and usability for SSO users.
A main difference pertains to the collection of user data APIs made available by each IdP. In many instances, the APIs are based on other services offered by the IdP, for example, a calendar API for Google Calendar, or a public profile API for Facebook Profile.
Here, we focus more on IdP platform design differences related to privacy and usability.

The granularity and customisation of SSO permissions vary considerably across different IdP platforms.
For instance, the Facebook Login platform~\cite{FacebookLogin} offers the option to opt-out of most user data APIs requested by RPs.
When users login with Facebook, they are provided a consent dialog that allows the user to deselect a subset of RP requested permissions and proceed with the login---although the opt-out option is nested in a secondary UI triggered from the consent dialog.
Some basic permissions such as Facebook profile picture and profile name cannot be opted-out (regardless of whether the RP has requested those permissions).
Google's consent dialog~\cite{SignInWithGoogle} instead uses an all or nothing permissions model where users are forced to either approve all or reject all of the RP requested permissions.

For a small subset of user data APIs, Facebook allows users to grant fine-grained permissions~\cite{FacebookLogin}.
For example, if an RP requests the permission to manage user-administered Facebook pages on the user's behalf, the user can choose to grant the RP access to only a certain subset of all the pages they administer.
But such fine-grained controls are not offered for many sensitive Facebook permissions (such as personal photos), and are completely absent in other IdP platforms including Google.

\subsubsection{\textbf{Pseudonymisation of personal information}}
\label{sec.background.idp.pseudonymousemail}
Apple's SSO platform~\cite{Apple2023PrivateEmailRelay} offers its users the option to login to an RP site using a custom name and a pseudonymous email address.
Using this privacy feature, email messages between a user and an RP are transferred through Apple's email relay service without revealing the user's actual email address to the RP.
This enables the RP to communicate with the user while the IdP restricts the RP's ability to track or sell the user's email address.
This feature might be preferable to privacy-conscious users with an active Apple account, but requires trusting Apple with any personal information that might be revealed in emails from/to the RP.
Other IdPs including Google and Facebook do not support similar privacy options.

\subsubsection{\textbf{RP App Reviews by IdPs}}
Many IdPs categorise their user data APIs into a basic API set for user profile data (e.g., name, email address) that might be less private and/or necessary to provision RP services, and a separate privacy-sensitive API set for more personal user data such as user photos and email messages.
RP sites that depend on these privacy-sensitive APIs are typically required to go through app reviews by IdPs before requesting access to user data.

\textbf{\textit{Google SSO.}}
Google's app verification~\cite{google2023OAuthAppVerification} is divided into three levels of assessments depending on the user data APIs a given RP app wants to access.
All RPs undergo a basic ``brand verification'' process which involves IdP checks of the information provided by the RP for inclusion in the IdP's consent dialog (e.g., RP domain name, RP privacy policy URL).
RPs that request privacy-sensitive APIs (e.g., Google Calendar API) must accept Google's User Data Policy which requires the RP to agree to certain privacy practices such as requesting the minimum set of permissions necessary for services.
RPs that request further restricted APIs (e.g., Gmail APIs) must additionally go through a \textit{security assessment} based on Google's App Defense Alliance\footnote{\url{https://appdefensealliance.dev/casa}} testing program.
This program was launched in 2019 to detect security threats in Google Play Store apps, and has since been expanded to cover third-party services including RP sites that request certain restricted user data APIs~\cite{google2022ExpandingADA}.

\textbf{\textit{Facebook SSO.}}
If the RP requested permissions are beyond the basic public profile or email address APIs,
Facebook's App review~\cite{Facebook2023AppReview} requires RP developers to get approval before accessing the APIs.
This review involves a basic brand verification check for RP's privacy policy URL (similar to Google above).
The verification is also supposed to check whether the RP site uses the requested user data APIs to provision one or more of its services.
During the review process, developers are asked to explain why their app requires access to the requested APIs, and to submit video demos of app features that utilise the accessed data.
This aims to limit RPs from accessing user data unless needed by some legitimate functionality, but RPs with the primary aim to track users might introduce superficial functions exclusively designed to pass the IdP review processes.
\newline

\noindent
These IdP app reviews aim to limit misuse of user data by RPs for purposes other than providing RP services, for example, by checking if requested user data is necessary for any relevant RP app functionality.
When evaluating the necessity of the requested user data, the scope of the IdP reviews are limited to only that IdP's login option and the RP functionality directly related to its user data APIs.
Such evaluation could miss RPs that request varying amounts of user data across different SSO login options~\cite{morkonda2021empirical}, meaning that the user data APIs claimed as necessary with one IdP login choice might not be claimed as necessary with other IdP login options.

\section{Methodology}
\label{sec.evaluation.survey}
We conducted an anonymous survey using a between-subjects design to investigate how users choose from a list of different SSO and non-SSO login options based on the current SSO login interface (RQ1).
This study also investigates the effect on login decisions of an augmented SSO workflow where comparative information about IdP permissions is displayed before prompting users for a login choice (RQ2).
We implemented the survey using the Qualtrics\footnote{\url{https://www.qualtrics.com}} online survey platform and collected responses in Jan-Feb 2023.

\begin{figure}[tb]
    \centering
    \frame{\includegraphics[width=0.68\columnwidth]{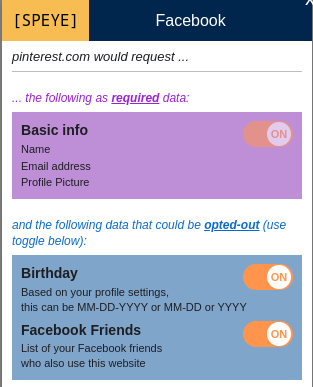}}
    \caption{SPEye~\cite{morkonda2022ssoprivateeye} interface showing the permissions requested by an example RP site through Facebook SSO login.}
    \label{figSpeyePinterest}
\end{figure}

\subsection{Extracting SSO permissions from RP sites}
\label{sec.methodology.ssoprivateeye}
In parallel work, we designed and built a Chrome browser extension called SSOPrivateEye (SPEye)\footnote{Design and implementation details of the tool are given in a companion paper~\cite{morkonda2022ssoprivateeye}.}, to augment SSO workflow to show users privacy-related information about the permissions requested by an RP through individual SSO login options before users login.
SPEye~\cite{morkonda2022ssoprivateeye} scans an RP website for SSO login options and extracts the permissions (from the standard OAuth \texttt{scope} parameter described in Sec.~\ref{sec.background.websso}) the RP would request from users through each of the available SSO login options.
Figure~\ref{figSpeyePinterest} shows SPEye's UI displaying the permissions requested by an example RP site with Facebook login.
The permissions are highlighted in two color-coded blocks to differentiate between mandatory permissions required for login with a given IdP and permissions that can be opted-out (as designated by the individual IdPs).

Standard OAuth~\cite{oauth2rfc} requires RPs to specify the permissions in the initial authorization requests to the IdPs, thus information about RP permissions is available to automated tools such as SPEye and can be displayed before users login, even without the cooperation of RPs.

The present user study complements user tools such as SPEye by testing the effect of comparative permission information on users' login decisions.
We used screenshots of SPEye's output on different RPs to show comparative privacy information about RP permission requests for RQ2, as described next. The user study was not intended to assess the usability of SPEye specifically, but rather to explore the impact of displaying this type of information.

\subsection{Study overview}
\label{sec.methodology.studyoverview}
The survey was built around the login options in four popular RP sites (\url{Airbnb.ca}, \url{CBC.ca}, \url{NYTimes.com}, and \url{Rakuten.com}).
Our intention was not to assess these specific RPs but to represent a variety of requested login permissions and a range of privacy choices.
For example, \url{Rakuten.com} requested more privacy-intrusive permissions with Google SSO compared to other options, whereas \url{CBC.ca} requested most permissions with Facebook SSO. All four RPs offered exactly three SSO login options (Google, Facebook, and Apple) and a non-SSO login option using an email address; \url{Airbnb.ca} users could also log in using their phone number.
For each RP site, we collected and displayed screenshots of SPEye's output (Sec.~\ref{sec.methodology.ssoprivateeye}) for each IdP login option to enable participants to compare the permissions requested by the RP.

We designed the survey to show participants the current login interface of an RP and prompt for an initial login choice (Pre-SPEye) along with an explanation for their choice. Next, participants saw the augmented view provided by the SPEye interfaces facilitating comparison of permissions and they again entered a login choice (Post-SPEye) and explanation. To preserve the validity of responses, participants could not navigate back to earlier survey pages. 

To avoid priming effects, each participant was shown only one of the four RP sites. We configured our survey recruitment to allocate exactly 50 participants per RP, for a total of 200 participants.
The average completion time for the survey was 6 minutes and 23 seconds, and participants were paid \textsterling 1.50.

The survey (shown in Appendix~\ref{sec.app.survey}) was structured as follows: 

\begin{enumerate}[left=0pt .. \parindent, nosep]
\item \,Basic demographic questions.
\item \,Pre-SPEye login: we showed each participant a screenshot of the usual login prompt of their assigned RP and prompted for their preferred login option (among both SSO and non-SSO) using a multiple choice question, followed by an optional open-ended question for the rationale behind their choice.
By showing the usual login prompt of the RP, we prompt participants to choose based on the information available to users at the time of login.
\item \,Post-SPEye login: we presented screenshots of SPEye's output on the individual IdP login pages, and again prompted the participant for a login choice, with an optional open-ended question for their rationale.
This time the login options in the question included additional privacy choices made apparent by SPEye's output (e.g., Apple's email relay service described in Sec.~\ref{sec.background.idp.pseudonymousemail}). In essence, we explore what happens in an augmented workflow where additional privacy-related information is made visible and easily comparable.
\item \,We asked 5-point Likert scale questions with statements related to their login choice and privacy preferences.
\item \,We asked participants to indicate if they had an account with any of the popular IdPs (including Apple, Facebook, and Google), and if so, how often they used applications and/or services (related to both SSO and non-SSO usage) connected to their IdP accounts.
\end{enumerate}

\subsection{Participants}
We recruited 202 participants in the US and Canada through the Prolific\footnote{\url{https://www.prolific.co/}} recruitment platform.
Responses from two of the participants were excluded from the analysis as they completed the survey in less than two minutes and selected the same option for each Likert scale question, leaving us with 200 valid responses.
There were 97 participants who self-identified as men, 97 as women, 3 as non-binary, one fluid, one demigirl, and one preferred not to answer.
The self-reported ages (excluding a participant who preferred not to answer) ranged from 19 to 80 ($M = 37.2$, $SD = 13.9$).
The majority of participants were university graduates or students currently enrolled in a university degree; 43\% with a Bachelor's degree, and 21\% with a Master's or PhD degrees.

Among the 198 participants who responded, 48\% said they log into 3-5 different sites (not pertaining to any specific login method) on a typical day and another 30\% reported logging into 6-8 sites per day.
The majority of participants (90\% of 199) had previously used web SSO to log into at least one site, either with a work account (e.g., office email) or a personal web account. 
When asked about web SSO use outside work (197 responses), 59\% reported using SSO on 1-5 sites outside work and 15\% reported not using SSO outside work. We report additional details on demographics, web login, and web SSO accounts used by participants in Table~\ref{tableDemographics}.

\begingroup
\begin{table}[tb]
    \centering
    \caption{Participant demographics and web SSO usage.}
    \label{tableDemographics}
    \begin{tabular}{llr}
        \toprule
        \textbf{Demographics}\\
        \midrule
        Gender ($N=199$) & Man & 49\%\\
        & Woman & 49\%\\
        & Other & 2\%\\
        \midrule
        Age ($N=199$) & 18-19 & 2\%\\
        & 20-29 & 33\%\\
        & 30-39 & 28\%\\
        & 40-49 & 18\%\\
        & 50-59 & 11\%\\
        & 60+ & 8\%\\
        \midrule
        Education ($N=197$) & High school & 18\%\\
        & College & 18\%\\
        & Bachelors & 43\%\\
        & Advanced degree & 21\%\\
        \midrule
        \textbf{Web login and SSO usage}\\
        \midrule
        Daily web logins ($N=198$) & 0-2 & 4\%\\
        & 3-5 & 48\%\\
        & 6-8 & 30\%\\
        & 9-11 & 10\%\\
        & 12+ & 8\%\\
        \midrule
        Used SSO ($N=199$) & Yes & 90\%\\
        & No & 7\%\\
        & Unsure & 3\%\\
        \midrule
        Non-work accounts & 0 & 15\%\\
        linked with SSO ($N=197$) & 1-5 & 59\%\\
        & 6-10 & 15\%\\
        & 11-15 & 6\%\\
        & 16-20 & 1\%\\
        & 21+ & 4\%\\
        \bottomrule
    \end{tabular}
\end{table}
\endgroup

\subsubsection*{\textbf{IdP accounts}}
\label{sec.participants.idpaccounts}
Table~\ref{tableWebAccountUseFrequency} presents a summary of participants' use of apps and services offered by Google, Facebook, and Apple, which are IdPs widely supported by RP sites for web SSO login.
Of the 198 responses, 96\% had a Google account, 80\% had a Facebook account, and 62\% had an Apple account.
Most participants reported using Google apps/services frequently, with 78\% who used a Google service at least once a day, and 12\% who used once per week.
In contrast, less than half of participants reported using their Facebook account (35\%) or their Apple account (26\%) on a daily basis.
The remaining participants reported using Facebook or Apple infrequently; 26\% of the participants for Facebook and 25\% for Apple said they used their accounts once a month or less.

Some of these self-reported values may be underestimated because they involve cases where a participant reported using their account (such as Apple) infrequently but in practice uses their account regularly by using a device (e.g., iOS smartphone) associated with their IdP account.

We compared the use of IdP accounts among participants assigned to each RP site.
A similar number of participants reported frequent (daily or weekly) use of a Google account (ranging from $43-46$ for the four RPs) and of an Apple account (ranging from $17-20$).
There was greater variation across the RPs for frequent use of Facebook accounts (ranging from 20 at Rakuten.com to 34 at Airbnb.ca).
This could possibly affect the number of participants per RP site choosing or avoiding Facebook, but we expect the impact to be low given the relatively small fraction of the total study participants who chose Facebook (pre-SPEye: 11 participants and post-SPEye: 11 participants).

\begingroup
\begin{table}[tb]
    \centering
    \caption{Participants' level of interaction with applications and/or services related to non-work accounts, based on a sample size of 198 participants who responded to the demographics questions.}
    \label{tableWebAccountUseFrequency}
    \begin{tabular}{lrrr}
        \toprule
        \textbf{Frequency of account use} & \textbf{Google} & \textbf{Facebook} & \textbf{Apple} \\
        \midrule
        at least once a day & 78\% & 35\% & 26\%\\
        at least once a week & 12\% & 19\% & 11\%\\
        at least once a month & 4\% & 13\% & 14\%\\
        at least once a year & 1\% & 10\% & 10\%\\
        never & 1\% & 3\% & 1\%\\
        \midrule
        no account & 4\% & 20\% & 38\%\\
        \bottomrule
    \end{tabular}
\end{table}
\endgroup

\subsection{Ethical Considerations}
This low-risk study was reviewed and cleared by our university's Research Ethics Board. To protect privacy, we collected no information about participants' login credentials and we did not ask participants to directly interact with our test sites.  Identifiable information was limited to Prolific's generated participant identifier, which we stripped from the data set before starting analysis. 

\subsection{Limitations}
We note that our data is collected from participants' self-reported responses on login decisions and related reasons.
These responses might not fully align with actual user behaviour in practice.
Participants reported their login decisions based on one of the four RP sites used in this study. 
These responses may not reflect participants' preferences on other RP sites, e.g., on a different category of site, or on RPs with a different set of login options. 
Our sample of RPs represents a small variety of sites that support web SSO systems.
We recruited participants in the US and Canada using the Prolific platform; these participants, by virtue of having signed up for online survey platforms, may differ (e.g., by possibly being more tech-savvy) from the general population.

\section{Login decision results}
\label{sec.evaluation.survey.quantitative}
\begin{figure}[tb]
    \centering
    \includegraphics[width=1.09\columnwidth]{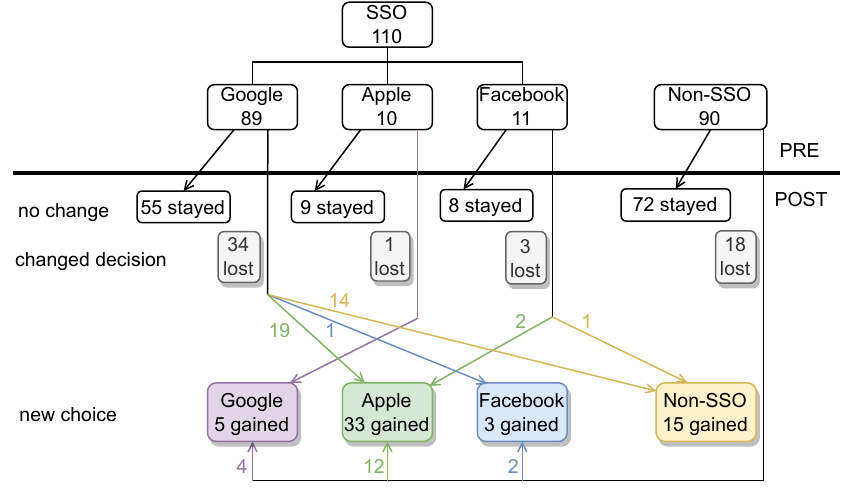}
    \caption{Distribution of the pre- and post-SPEye login decisions. The decisions that changed (coloured cells) are further broken out to show the new login choices. Non-SSO includes password-based login with email address or phone number.}
    \label{figPrePostChoices}
\end{figure}

Figure~\ref{figPrePostChoices} summarises the participants' login decisions before and after viewing SPEye comparative information about IdP permissions.
Where participants changed their login decisions, we observed a tendency to move towards more privacy-friendly login options.

\subsection{\textbf{Pre-SPEye login decisions}}
\label{sec.quantitative.prespeye}
$55\%$ of participants ($n=110$) preferred an SSO login option as their pre-SPEye login choice (i.e., asked on a separate page before we showed SPEye's screenshots).
Among the SSO users, Google login was by far more popular ($81\%; n=89$) than Facebook ($10\%; n=11$) or Apple ($9\%; n=10$) login.
Common reasons given for selecting Google included convenience because participants were already using other related services (such as Gmail) that results in them remaining logged in with Google, and the convenience of linking RP accounts to their primary web account.
This is consistent with the large number of participants ($76\%$ of 200) who indicated using Google services daily; versus $31\%$ for Facebook, and $21\%$ for Apple.

Non-SSO login options (in our study, password-based) were selected as an initial choice by 90 participants ($45\%$), and common reasons given were privacy concerns related to SSO linking of multiple RP accounts with the IdP.
Additionally, some participants perceived security benefits from having separate accounts to reduce the risk of one compromised account impacting another.
Besides releasing personal data to RPs, some non-SSO users noted concerns about IdP tracking of RP site visits.
Sec.~\ref{sec.results.qualitative.prespeye} provides further details.

\begin{table*}[tb]
\centering
\caption{Number of participants choosing each SSO login before and after viewing comparative information about IdP permissions. Choices for each RP are ranked from least (1) to most (6) privacy-friendly depending on the user data revealed to the RP (e.g., the \faFacebookSquare\textsuperscript{\textdagger} option on Airbnb is ranked higher than {\faFacebookSquare} as it opts out of the birthday permission). 
Cells with 2 icons mean those options are privacy-equivalent, and the corresponding count is the sum for both options.}
\begin{threeparttable}
\begin{tabular}{r|lrr|lrr|lrr|lrr|rr}
\toprule
& \multicolumn{3}{c|}{Airbnb} & \multicolumn{3}{c|}{CBC} & \multicolumn{3}{c|}{NYTimes} & \multicolumn{3}{c|}{Rakuten} & \multicolumn{2}{c}{Total} \\
Rank & login & pre & post & login & pre & post & 
login & pre & post & login & pre & post & pre & post \\
\midrule
1& \faFacebookSquare & $4$ & $2$ & \faFacebookSquare & 1 & 0 & - & - & - & \faGoogle & $25$ & $10$ & $30$ & $12$ \\

2& - & - & - & \faGoogle & 19 & 16 & - & - & - & - & - & - & $19$ & $16$ \\

3& \faGoogle, \faFacebookSquare\textsuperscript{\textdagger} & 24 & 23 & \faFacebookSquare\textsuperscript{\textdagger} & - & 2 & \faGoogle, \faFacebookSquare & 24 & 15 & \faFacebookSquare & 3 & 3 & $51$ & $43$ \\

4& \faApple & 3 & 3 & \faApple & 2 & 5 & \faApple & 3 & 1 & \faApple & 2 & 3 & $10$ & $12$ \\

5& \faEnvelopeO & 19 & 18 & \faEnvelopeO & 28 & 24 & \faEnvelopeO & 23 & 20 & \faEnvelopeO & 20 & 25 & $90$ & $87$ \\

6& \faApple$^\ddagger$ & - & 4 & \faApple$^\ddagger$ & - & 3 & \faApple$^\ddagger$ & - & 14 & \faApple$^\ddagger$ & - & 9 & - & 30 \\
\midrule
\multicolumn{13}{l}{Wilcoxon Signed-rank test (*statistically significant, i.e., $p<0.05$)} \\
& \multicolumn{3}{c|}{$p=0.0593$} & \multicolumn{3}{c|}{$p=0.1919$} & \multicolumn{3}{c|}{\textbf{\textit{p = 0.0004}}*} & \multicolumn{3}{c|}{\textbf{\textit{p = 0.0006}}*} \\
\bottomrule
\end{tabular}
\begin{tablenotes}
    \item dash (-) means option unavailable, rather than that 0 selected it.
    \item \faEnvelopeO: Non-SSO login option (email address or phone number)
    \item \faApple$^\ddagger$: Apple login option with \textbf{pseudonymised data}
    \item \faFacebookSquare\textsuperscript{\textdagger}: Facebook login option with \textbf{permissions opted out}
\end{tablenotes}
\end{threeparttable}
\label{tablePreferenceChange}
\end{table*}

\subsection{\textbf{Post-SPEye login decisions}}
\label{sec.quantitative.postspeye}
Overall, 28\% of participants ($n=56$) decided to change their login choice after viewing the comparative IdP permissions.
Of these participants, 23 moved from one SSO login option to another SSO option, 18 moved from non-SSO to SSO login options, and 15 moved from SSO to a non-SSO login option. Fig.~\ref{figPrePostChoices} shows these changes.

Among the 89 participants who initially chose Google login (the most popular pre-SPEye SSO login choice), 34 changed their preference: 19 changed to Apple login and 14 changed to password-based login.
The primary reason given for this change was the availability of a more privacy-friendly option.
Apple login was more privacy-friendly because it offered a privacy option to pseudonymise the user's name and email address (Section~\ref{sec.background.idp.pseudonymousemail}) and to use SSO without revealing real (personal) information to the RP.
In total, 42 of the 200 participants preferred to login with Apple after viewing SPEye's screenshots; and 30 of these participants said they would use Apple's pseudonymisation privacy feature.

During our analysis, we ordered the login choices per RP using a 6-point scale based on apparent privacy-friendliness, given the data released to the RP\footnote{This ordering was not visible to participants taking the survey}. 
For example, we rank the five login choices on Rakuten.com from 1 (least privacy-friendly: default Google login option requesting access to personal email messages) to 6 (most privacy-friendly: Apple login option with pseudonymised data).
Within an RP, we assigned the same rank to login options with comparable privacy---e.g., the Google and Facebook login options on NYTimes.com which requested basic information (name, email address, and profile picture) are both ranked as 3 (privacy-neutral).
Table~\ref{tablePreferenceChange} shows the privacy ranking for each RP and the distribution of participants' pre- and post-SPEye login decisions.

We conducted a Wilcoxon Signed-rank test comparing the pre- and post-SPEye login decisions for each RP, and found statistically significant differences for two RPs (NYTimes and Rakuten), where the post-SPEye decisions were more privacy-friendly, as reported in Table~\ref{tablePreferenceChange}.
We found no statistically significant differences for the Airbnb and CBC RPs.
However, among the participants who changed (CBC: $n=12$, Airbnb: $n=13$), most of them moved to a more privacy-friendly option (CBC: $n=8$ and Airbnb: $n=10$).
Furthermore, the majority of the pre-SPEye login choices for Airbnb and CBC RPs were already privacy-friendly (i.e., the least privacy-friendly choice, Facebook login, was initially chosen by only four and one participants respectively).
Many of the participants who selected Google login for CBC (ranked as second least privacy-friendly choice) said in their open-ended responses that they were comfortable with their pre-SPEye decision as they considered the requested permissions (i.e., name, email address, language preference, profile picture, and public profile) acceptable.
Sec.~\ref{sec.results.qualitative.postspeye} provides further details on participants' rationales for login decision changes.

\begin{figure}[tb]
    \centering
    \includegraphics[width=0.97\columnwidth]{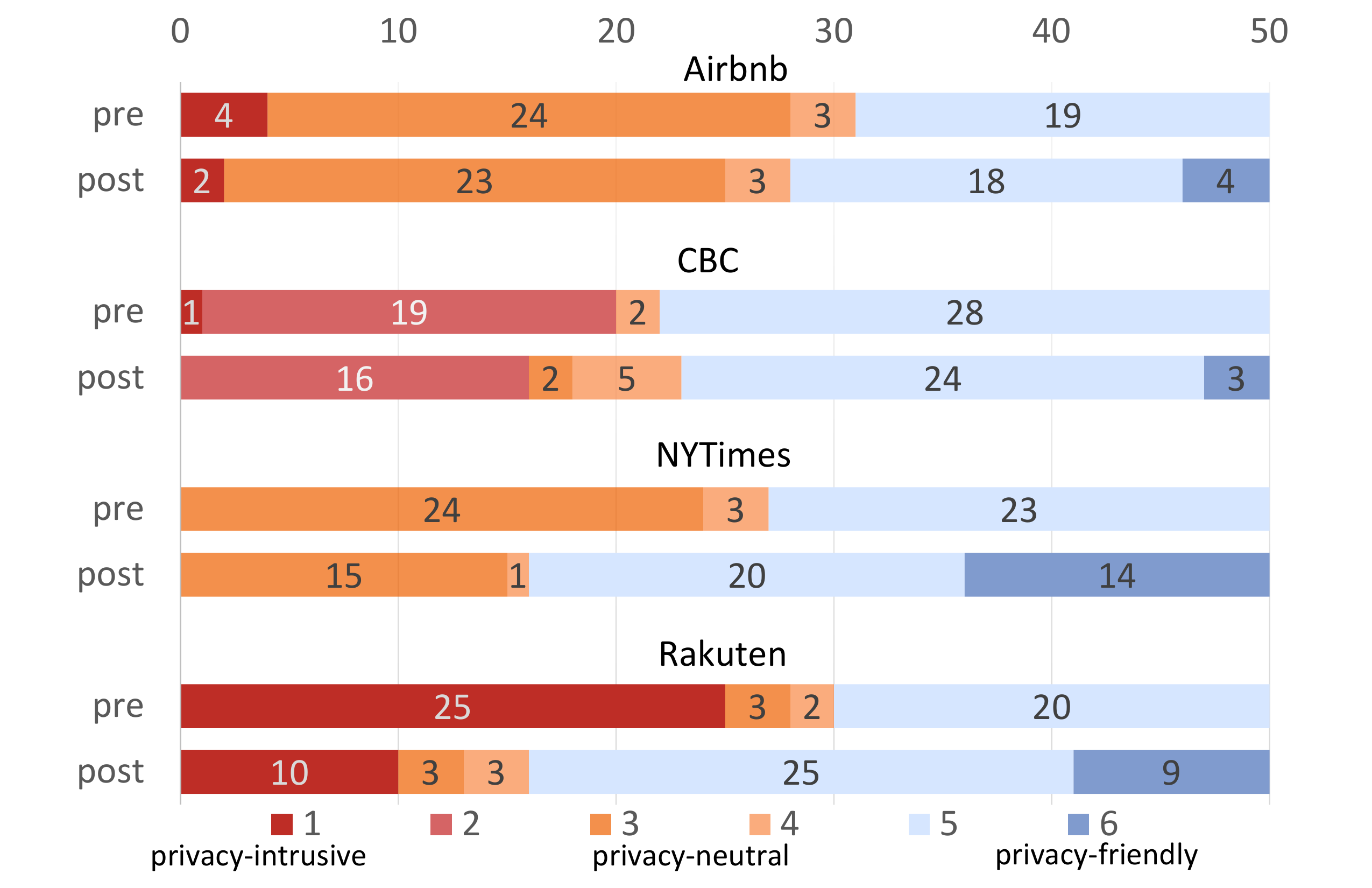}
    \caption{Participants' pre- and post-SPEye login decisions shown per RP. Login choices for each RP are ordered from privacy-intrusive (1) to privacy-friendly (6) to show shifts in the privacy of login decisions after viewing IdP permissions.}
    \label{figLoginChoices}
\end{figure}

\subsection*{Summary of effects of comparative information}
Fig.~\ref{figLoginChoices} shows the privacy of pre-SPEye login decisions and highlights the shift in participants' post-SPEye decisions towards more privacy-friendly login options.
Overall, we found 48 of 56 participants who changed their login decisions (i.e., 86\% of those changing) moved to a more privacy-friendly login option after viewing the IdP permissions.
Among the remaining 72\% of participants ($n=144$) who did not change, many of them had chosen login options that were privacy-neutral (ranked 3-4) or privacy-friendly (ranked 5-6) already; e.g., 81 of these 144 participants chose a non-SSO login option (ranked 5, privacy-friendly) or the Apple login option (ranked 4, privacy-neutral), as implied by Fig.~\ref{figPrePostChoices}.

These results suggest to us that many users will make more privacy-informed decisions when provided with comparative information about the differences between IdPs with respect to requested IdP permissions.

\section{Factors influencing login decisions}
We now summarise the qualitative analysis of the reasons given by participants for their login decisions. We outline the factors that influenced their pre- and post-SPEye decisions, and highlight the effects of showing comparative information about IdP permissions.

\subsection{\textbf{Coding methodology}}
We collected 333 open-ended responses relating to participants' reasons for their pre-SPEye ($n=176$) and post-SPEye ($n=157$) login decisions.
These responses were coded using the thematic analysis method by Braun and Clarke~\cite{braun2012thematic}.
Inductive coding is a process where the responses are carefully reviewed and conceptualised into an initial set of codes (codebook) which are iteratively revised as new codes are recognised.

The lead researcher first reviewed the 176 pre-SPEye responses to develop a set of 14 codes, then added 7 codes from reviewing the 157 post-SPEye responses, forming an initial codebook containing 21 codes in total.
The research team met several times to iteratively review and revise the codebook and discuss the classification of individual excerpts, resulting in a final set of 17 codes including the pre- ($n=13$) and post-SPEye ($n=4$) codes.
All three members of the research team agreed with the final categorisation of excerpts into codes.
We further grouped the 17 codes into 7 themes (6 pre- and one post-SPEye) based on common topics that emerged during our analysis.
Table~\ref{tableCodebook} lists these individual codes and themes.

The results discussed next are split into pre- and post-SPEye themes along with sample excerpts for the individual codes.
For each theme, we include the number of login decisions that were influenced by the codes to differentiate more popular themes.
When a participant's response indicated multiple reasons for their login decision (e.g., privacy and security), we counted the response in each of the themes.
Thus, a single participant's pre- (or post-SPEye) response might be split into different pre- (or post-SPEye) excerpts and included in the counts for multiple factors.
In addition, responding to the open-ended questions was optional so our analysis is based on participants who provided valid responses indicating reasons for their login decisions.

\begingroup
\begin{table*}[tb]
    \centering
    \footnotesize
    \caption{Themes that emerged during our analysis of the participant responses for their pre- and post-SPEye login decisions. The right two columns represent codes relating to participants who chose SSO and non-SSO login options. Post-SPEye codes marked with a dagger $\dagger$ denote those derived from responses after participants viewed the comparative information about IdP permissions.}
    \label{tableCodebook}
    \begin{tabular}{llll}
        \toprule
        \multicolumn{2}{l}{\textbf{Theme}} & \textbf{SSO login codes} & \textbf{Non-SSO login codes} \\
        \midrule

        1. & Usability benefits & SSO usability better & Non-SSO usability better\\
        & and obstacles\\
        \midrule

        2. & Inertia/habits & Use of primary account & Use of existing non-SSO account \\
        & & Use of existing RP/IdP account\\
        \midrule
        
        3. & Perceived security & Perceived SSO security benefits & Perceived non-SSO security benefits\\
        \midrule

        4. & Privacy-related & Self-managed privacy strategy & Non-SSO privacy better\\
        & motivations & Privacy is not a major concern\\
        & & $\dagger$ Provides better privacy\\
        \midrule

        5. & Linked account & & Concerns related to linking accounts\\
        & concerns\\
        \midrule

        6. & Trust perceptions & Perceived trust/reputation in the IdP & Concerns related to trusting SSO entities and the overall system\\
        \midrule

        7. & Privacy-usability & $\dagger$ Good compromise between privacy and usability & $\dagger$ Privacy concerns but lack existing privacy-friendly IdP account\\
        & tradeoffs & $\dagger$ Usability-privacy conflict but chose usability\\
        \bottomrule
    \end{tabular}
\end{table*}
\endgroup

\subsection{\textbf{Pre-SPEye factors}}
\label{sec.results.qualitative.prespeye}
The pre-SPEye login decisions were primarily based on \textit{usability} preferences and \textit{inertia} factors as more than 50\% of pre-SPEye excerpts were related to these two themes.
\textit{Security} and \textit{privacy}-related factors were mentioned in around 25\% of the pre-SPEye excerpts, and the remaining two themes (\textit{linked account concerns} and \textit{trust perceptions}) emerged in 15\% of the pre-SPEye excerpts.
\newline

\noindent
\textbf{1. Usability benefits and obstacles.}
We define usability in the web SSO context as relating to convenience in storing/retrieving passwords and possibly related to other services and/or devices used by the user.
Where login also involves release of personal information to RP sites, usability also relates to convenience in managing the personal data in the IdP account.

Usability was a dominant factor in pre-SPEye login decisions, especially among participants ($n=53$) who chose SSO login options.
These participants' responses indicated that SSO was convenient because they did not have to create or remember new passwords on individual RP sites.
SSO was also preferred for fast login because participants reported remaining signed in to their IdP account on a related website or device which they used regularly.

\begin{quote}
``\textit{I'm usually logged into Gmail so it's easier to use my Google account [to login].'' (P186)}
\end{quote}

Conversely, a smaller group of participants ($n=15$) found non-SSO login options simple and easy to use.
These participants mentioned using tools such as password managers and browser autofills which offer users comparable login experiences as SSO login options, sometimes with perceived added benefits of having separate accounts.
For example, P21 felt they had better control over their personal information with non-SSO accounts.
Using non-SSO accounts, users can choose what personal information they share with the sites they log into, as opposed to SSO accounts where the information might be linked from the IdP account.

\begin{quote}
    \textit{``I prefer having a separate account for sites as it allows me to better customize my privacy and profile settings [...]'' (P21)}
    \newline
\end{quote}

Other benefits mentioned by participants included easier recollection of previously chosen logins, e.g., by always choosing login with email and password for web logins across different web sites.

The perceived usability of SSO and non-SSO login options depended on which features were most valued by individuals, and their context of login use with respect to other web accounts.
Participants primarily associated SSO with better usability as it removed the need to create/remember extra passwords and enabled fast login using IdP accounts they were already logged into at a frequently used IdP service.
Comparatively fewer participants associated non-SSO login with better usability, though participants felt better control over their data which justified their non-SSO decision.
\newline

\noindent\textbf{2. Inertia/habits.}
Inertia in software systems is defined as user tendency to persist with the use of an existing system in the presence of alternate options~\cite{polites2012shackled}.
We regard inertia in the context of web login as the tendency of a user to persist with an existing login method/habit when other alternative login options are available.

Inertia was the second most dominant theme as participants preferred login options they were already habituated with from using across different sites.
Similar to the usability theme, most of the participants in this theme chose SSO login options ($n=48$) compared to non-SSO options ($n=13$).
Participants primarily preferred login options associated with their primary IdP accounts because they were most familiar with these accounts and were often signed in with other related IdP apps (e.g., Gmail, Safari browser).

\begin{quote}
``\textit{I have used Google for many years so its [sic] my go to when logging onto different sites.'' (P42)}
\end{quote}

Participants who did not have a strong preference for how to login were influenced to choose SSO login because they preferred to use their primary email address, with or without SSO. For example, P167 chose SSO login because their primary email address was associated with an IdP:
\begin{quote}
``\textit{[Google login is] faster and even if I was to sign up for an account, [I] would use my [G]mail for it anyways.'' (P167)}
\end{quote}

Some participants mentioned other benefits with primary accounts such as account longevity (i.e., continued use of primary account in future), and easier SSO account management from connecting different accounts to a central primary account.

Existing login habits influenced many participants' login decisions, especially among those who preferred SSO login options using primary accounts.
This preference aligns with web SSO design which connects different web login accounts to a primary IdP account.
We found that a significantly larger number of participants were habituated with SSO login options (78\% of those observed to be choosing based on inertia) compared to non-SSO login options.
This result contrasts findings in a 2011 SSO study~\cite{sun2011makes} where users generally preferred non-SSO login and were habituated to login using email address and password.
Our results suggest that many users have become more habituated with web SSO login over the last decade, possibly due to the increasing presence of SSO login options on RP sites, and popular IdPs such as Google and Apple.
\newline

\noindent\textbf{3. Perceived security.}
In our context, perceived security involves perceptions about the security of web login options including the security of login credentials, and in the case of SSO, any personal information requested through IdP login accounts.

Security perceptions influenced pre-SPEye login decisions among 25 participants. 
Like usability, participants articulated security arguments for and against both SSO and non-SSO login options, depending on their perspectives.
However, unlike themes 1 and 2 above, the majority of participants in this theme prioritizing security (19 of 25) chose a non-SSO login option.
Participants felt that using unique passwords offered better security than SSO options, possibly because they used password generators to create and manage passwords that are difficult to guess, or due to misconceptions about SSO access tokens which are less vulnerable to guessing attacks.
Another common perception among participants was that non-SSO accounts offered better security because a security breach of one account would not put at risk their other accounts.
\begin{quote}
    \textit{``Separate logins are more secure in the event of data breach. You don't need someone to have access to all of your accounts because they have access to your [G]oogle [IdP] account.'' (P11)}
\end{quote}

\begin{quote}
    \textit{``[...] It is more secure to use email and password combination for sign in, if a unique password is used for each website.'' (P199)}
\end{quote}

Although SSO protocols aim to improve security by replacing user-chosen passwords with unique access tokens, concerns about a single point of failure motivated participants to choose non-SSO login options. These concerns are not completely baseless, though it is unclear whether participants had knowledge of specific risks or whether they were generally uneasy about using SSO.  
A vulnerability in IdP code could be exploited by attackers to compromise access tokens; for example, in 2018 attackers exploited a vulnerability in Facebook's SSO software to gain access to 50 million user accounts, including access tokens linked to every compromised account~\cite{fb2018ssobreach}.
\newline

\noindent\textbf{4. Privacy-related motivations.}
Privacy in the web login context relates to personal information entered or linked to login accounts, including data released from an IdP to an RP site during SSO login.

Privacy-related reasons motivated the pre-SPEye login decisions for 23 participants.
Most of these participants (14 of 23) chose non-SSO login options to share less personal information with RP sites, though participants had not viewed the permissions requested with the alternate SSO login options yet.
Separately, some participants avoided SSO due to concerns about the IdP's ability to gain knowledge about users' website visits (i.e., cross-site tracking) through SSO login events:

\begin{quote}
    \textit{``I am not comfortable with giving Google, Facebook, or Apple even more of my personal information than what they already have.'' (P125)}
\end{quote}

Other privacy-conscious participants ($n=6$) who selected SSO took steps towards privacy using IdP tools (e.g., permission opt-out options offered by some IdPs such as Facebook) or through strategies such as using throwaway accounts that contained less personal information or that they used infrequently.
The remaining participants ($n=3$) with comments relating to privacy explicitly mentioned that they had no major privacy concerns about using SSO on the presented RP website, which could be due to an indifference towards privacy, or due to other factors including personal opinions about the RP sites and IdP login options we showed them.

These findings show that privacy concerns (besides security perceptions as noted previously) are common reasons motivating non-SSO login choices.
While better privacy controls (e.g., opt-out of permissions requested by RPs) might incentivise a subset of privacy-conscious users to use SSO, other concerns including cross-site tracking by IdPs influence users to avoid SSO altogether.
\newline

\noindent\textbf{5. Linked account concerns.}
Besides security and privacy, a group of participants ($n=18$) had other concerns about linking accounts which motivated them to choose non-SSO login options.
These participants avoided SSO because they were generally uneasy about linking different accounts, though they did not elaborate on these concerns.
A specific concern mentioned by P29 related to the freshness of the personal information in their older accounts:

\begin{quote}
\textit{``I get nervous about connecting my other accounts to websites because a lot of the options are accounts that I made when I was 10 [years old] and that data might not reflect me anymore.'' (P29)}
\end{quote}

Although participants did not explicitly mention privacy or security, it is possible that some of the concerns about linking accounts were related to privacy or security, similar to many of the participants in the previous two themes.
\newline

\noindent\textbf{6. Trust perceptions.}
Trust perceptions in the web SSO context relate to the confidence users might have in one or more entities involved in the SSO login system including RPs, IdPs, and other components in the SSO ecosystem (e.g., the protocols, and software components in browsers and servers).
This confidence may include perceptions of an entity's intentions (perhaps malicious) and their competency or ability to protect user information; our study is unable to distinguish between these aspects of confidence (or ``trust'').

Trust perceptions influenced pre-SPEye login decisions among 13 participants, and these were mainly based on their perceived reputation of IdPs.
One particularly common perception among the participants who chose SSO login ($n=6$) was that they trusted Google more than Facebook or Apple:

\begin{quote}
    \textit{``Google probably shares my personal information less often that the other [login] options.'' (P89)}
\end{quote}

This selective trust of Google did not extend to the remaining participants who chose non-SSO login options ($n=7$).
These participants did not want the IdPs to know about their RP site visits as they did not trust any of the IdPs (Google, Facebook, or Apple).

Compared to the previous themes, trust perceptions had the smallest influence on pre-SPEye login decisions, as implied by the number of participants.
Where trust did influence login decisions, it was primarily based on participants' perception about the IdPs, suggesting that major IdPs (Google, Facebook, and Apple) are more influential than RPs in users' perception about web SSO systems.
\newline

\subsection{Post-SPEye factors}
\label{sec.results.qualitative.postspeye}
After explaining the reasoning for their pre-SPEye login decision, participants viewed a screenshot of an augmented SSO login interface displaying comparative information about the data requested of each IdP. They then had an opportunity to reconsider their decision in light of this new information.
We now compare the pre- and post-SPEye login rationales, and discuss how participants' decisions were affected (or not affected) by the augmented view.

Pre-SPEye login decisions were mainly based on usability and inertia factors, with privacy having only a small influence (motivating 12\% of decisions).
However, privacy-related motivations were among the dominant factors influencing post-SPEye login decisions among 37\% of participants ($n=73$).
Usability preferences and inertia were noted in fewer excerpts (compared to pre-SPEye), though these were the second and third most popular post-SPEye factors, respectively.
We also identified a new theme that emerged among participants who weighed both usability and privacy factors when making their choices.
We observed fewer post-SPEye excerpts associated with linked account concerns (8 from the previous 18) which included general concerns participants had about linking accounts.
This change is because many of the original 18 participants articulated their concerns about linked accounts (including privacy- and security-related) more clearly in the post-SPEye responses after viewing the requested permissions.

Security and trust perceptions were two factors that remained largely unchanged as participants expressed similar views on these factors,
and the number of participants influenced by these factors remained similar.
The lack of change in security perceptions is expected as we did not show any information about the security of login options.
Trust perceptions---which we observed in participants' responses to be based on perceived reputation of IdPs---influenced a similar number of participants as pre-SPEye.
We observed some SSO participants who mentioned trusting Apple login (besides others who mentioned trusting Google login) in their post-SPEye responses, possibly a result of an increase in the number of participants choosing Apple login (Sec.~\ref{sec.quantitative.postspeye}).
\newline

\noindent\textbf{1. Privacy-related motivations.}
Privacy-related reasons motivated the post-SPEye login decisions among 78 participants.

After viewing the permissions requested by the RP, participants expressed privacy concerns about disclosing personal information requested by RPs, including both basic (email address and profile picture) and sensitive personal data (e.g., email messages).
The availability of another login option with better privacy influenced many participants ($n=39$) to change their login decision to a more privacy-friendly option, with Apple SSO (pseudonymised data) being the most popular choice selected by 64\% of those changing.
For example, P169 who first chose Google SSO was concerned about revealing their email address to the RP and changed their preference to Apple SSO with the pseudonymisation option:

\begin{quote}
    \textit{``[...] I don't want [...] my email address being sold to other providers. I already get enough junk mail.'' (P169)}
\end{quote}

Among another group of participants ($n=27$), showing the requested IdP permissions reinforced their pre-SPEye login decisions which they considered privacy-friendly.
These participants did not change their login decisions, though they raised privacy concerns similar to the earlier group of participants.
This reinforcing effect was mostly observed among participants who chose non-SSO login options ($n=24$).
This aligns with our earlier finding among pre-SPEye decisions that most of participants motivated by privacy chose non-SSO login, suggesting that privacy-conscious users tend to choose non-SSO login options.
\newline

\noindent\textbf{2. Privacy-usability tradeoffs.}
A new theme emerged among a subset of participants ($n=17$) who considered both usability (most dominant pre-SPEye factor) and privacy preferences (most dominant post-SPEye factor) when making their post-SPEye decisions.

Many of these participants ($n=9$) weighed both privacy and usability of their login choices after being informed about the permissions requested by the RP.
While these participants were concerned about privacy, each chose an SSO login option and reported being comfortable with the personal information that would be released to the RP through their login choice.
For example, P10 changed their login decision from email and password to Google login because they considered the personal information being requested by CBC.ca through Google (name, email address, language preference, profile picture, public profile) as being publicly available.
Participants also mentioned approving the privacy-usability tradeoff for their specific assigned RP sites, suggesting that they might weigh these factors differently based on individual sites.

\begin{quote}
    \textit{``I don't mind [CBC.ca] having access to already fairly accessible information about me [...]'' (P10)}
\end{quote}

Some of the participants ($n=5$) valued the usability benefits of an SSO login option despite the option involved revealing extra personal information to the RP.
For these participants, although their preferred login choice did not align with their privacy preferences for the RP site, they favoured the usability benefits of their chosen SSO login option.
Another group of privacy-conscious participants appreciated the privacy of certain alternate IdP options, but they did not have existing accounts with those privacy-friendly IdP options.
These participants felt compelled to use non-SSO login due to their privacy concerns and lack of alternate IdP accounts.

\section{Related work}
In this section, we focus on prior work related to SSO user perceptions and user privacy, rather than work on tools and analyses related to SSO security.

Prior work on user perceptions related to SSO login systems have focused mainly on individual IdP platforms.
Balash et al.~\cite{balash2022security} investigated Google users' perceptions of third-parties that support Google login and access personal information from Google accounts.
Our study differs as it includes two other major IdPs besides Google and explores user perceptions of both SSO and non-SSO login options.
Bauer et al.~\cite{bauer2013comparison} considered three IdPs (Google, Facebook, and Google+) to study the effectiveness of the consent dialogs used by these IdPs.
Here, we focus on RP login prompts which are displayed before users see IdP consent dialogs.
Robinson and Bonneau~\cite{robinson2014cognitive} evaluated users' understanding of Facebook Connect's SSO permissions, focusing on users' interpretation of read and write permissions. 

Other studies that examined login decisions focused on individual factors including privacy and security. 
Egelman~\cite{egelman2013my} studied login decisions in sites with Facebook Connect, focusing on privacy and convenience aspects related to the platform.
We find similar privacy-usability tradeoffs, but in a wider context involving other IdP platforms and a variety of privacy choices.
Sun et al.~\cite{sun2011makes} explored factors influencing users to avoid OpenID SSO systems, finding single point of failure concerns and lack of trust in RP sites.
Our results show that security and trust perceptions affect not only influence users to avoid SSO, but also influence users to adopt SSO login options.
Besides login preferences, our study also investigated the effects of comparative privacy information on login decisions.

Increasing concerns about privacy issues have motivated researchers to build privacy-enhancing tools. 
Our companion paper~\cite{morkonda2022ssoprivateeye} presented SSOPrivateEye (SPEye), a Chrome extension designed to inform users of RP sites about the permissions requested through SSO login options.
Weinshel et al.~\cite{weinshel2019oh} built a browser extension to inform users about web trackers using privacy dashboards that visualise tracking activity.
Farooqi et al.~\cite{farooqi2020canarytrap} designed a privacy tool to detect third-party apps on Facebook that abused their access to users' personal information.

Empirical studies related to SSO privacy practices have focused on the personal information released to third-parties through SSO.
Morkonda et al.~\cite{morkonda2021empirical} collected privacy data on the permissions requested by RP sites through four IdP logins, and found many sites with significant privacy differences across the individual login options.
Järpehult et al.~\cite{jarpehult2022longitudinal} scanned RP sites in a longitudinal study over nine years to look for changes in the permissions requested by RPs.
Wang et al.~\cite{Na2011Third} analysed the privacy practices of third-party apps on Facebook, including the personal information requested by these apps.
Felt and Evans~\cite{felt2008privacy} examined the permissions in 150 Facebook apps and found that many apps requested data that was not essential for their services.

Many researchers have studied permissions and user privacy in mobile apps.
Cao et al.~\cite{cao2021large} collected data on users' privacy behaviour in Android app permissions, finding that users were more likely to grant permissions when informed about its use.
Wijesekera et al.~\cite{wijesekera2017feasibility} built an Android privacy classifier to infer privacy preferences of users based on previous app permission requests.
Felt et al.~\cite{felt2012how} compared different approaches in Android apps to request permissions and obtain consent from users.
Liu et al.~\cite{liu2016follow} proposed a privacy tool to manage permissions in Android apps according to users' privacy preferences.

\section{Discussion and Concluding Remarks}
Our study provided results showing participants' login decisions before and after viewing the IdP permissions requested by RP sites.
Inductive coding of open-ended responses providing rationale for their choices
uncovered factors that influenced these decisions, reflecting potential effects of showing the comparative IdP login permissions.
\newline

\noindent\textbf{Informed login decisions.}
After we showed the permissions requested by RPs, there was a shift in priorities (cf. Fig.~\ref{figLoginChoices}) towards more privacy-friendly login options; many participants expressed this was due to concerns about releasing personal information.
In some cases, we observed the reverse effect where users informed about the permissions made a privacy-usability tradeoff to share more information with RPs in exchange for the convenience of SSO.
We attribute these changes to enabling users to compare differences in IdP permissions across login options, and more visibly displaying privacy implications of different login options.

We encourage future work to explore how to inform users about the privacy implications and available privacy choices in SSO login systems.
In practice, informing users about differences in IdP permissions is challenging as current SSO UI designs focus on individual login options without providing sufficient context to compare the implications of different login choices.

Privacy tools such as SPEye~\cite{morkonda2022ssoprivateeye} demonstrate the feasibility of extracting privacy information about IdP permissions requested by RP sites and augmenting the current SSO workflow to show privacy comparisons.
We encourage IdPs and browser vendors to adopt UIs that enable comparison of permissions (such as elements of the SPEye UI used in our study), enabling privacy-informed user decisions.
\newline

\noindent\textbf{IdP tracking issues.}
Regarding factors influencing login decisions, we found that users motivated by privacy 
had a tendency to believe that choosing non-SSO login would offer privacy benefits including less information shared with RPs and avoiding cross-site tracking of their RP visits by IdPs.
However, the privacy mental models of users who chose non-SSO login were misguided with respect to avoiding cross-site tracking, because major IdPs such as Google and Facebook can still track non-SSO users, e.g., by tracking scripts such as Google's Doubleclick and Facebook's Pixel---which appear in a large number of web sites~\cite{englehardt2016online}.
We note that such scripts enable IdPs to track users independent of their use of SSO login.

We encourage future work to explore ways to increase awareness of broader tracking issues in the web SSO context where non-SSO users might not anticipate such tracking.
Although ad blockers such as Ublock Origin can stop tracking by Doubleclick and Pixel, these tools are not designed to stop tracking via IdP scripts related to web SSO toolkits~\cite{morkonda2022ssoprivateeye}.

Existing tools such as Tracking Transparency~\cite{weinshel2019oh} use privacy dashboards to inform users about third-party web tracking scripts including Doubleclick and Pixel.
Similar dashboards about IdP scripts can help inform SSO users about possible tracking by IdPs.
\newline

\noindent\textbf{Misunderstanding of release of personal information.}
Our study found that users who made login decisions based on trust factors (Sec.~\ref{sec.results.qualitative.prespeye}) were influenced by their perceptions of IdP reputation. This is despite the personal information requested depending primarily on the individual RPs, and being released to these RPs (which users may not have the same trust for).
This IdP-to-RP trust transference (users perceiving an RP as trustworthy based on the user's trust in the IdP, and the RP-IdP relationship) was more frequently observed with Google than Facebook or Apple (pre-SPEye), perhaps due to Google's popularity among participants.
This independent observation is consistent with findings of IdP-to-RP trust transference in a recent study specifically on Google SSO login~\cite{balash2022security}.
However, such trust transference contrasts earlier SSO studies (e.g.,~\cite{egelman2013my,sun2011makes}) that found that users based trust perceptions on RP reputation versus that of IdPs.

We encourage future work to further explore trust transference in SSO systems involving stakeholders such as RPs, IdPs, and browser vendors.
For example, the contrast described above could be due to a number of factors such as users being (intentionally or unintentionally) misguided by SSO UI designs about how SSO permissions work.
It might also be that IdPs that wish to keep users locked to their platforms have less incentive to improve transparency in permissions UIs.
Besides ineffective UIs, it is also possible that some users trust the IdPs (and therefore the RPs that offer these IdP options) that release less information to RPs, e.g., Apple SSO APIs currently only reveal name and email address.

Privacy ratings for SSO systems of popular RPs and IdPs (e.g., driven by privacy activists) may increase awareness of SSO privacy practices and encourage informed mental models of the release of personal information through SSO login.
Beyond SSO, Mozilla has published similar privacy ratings\footnote{\url{https://foundation.mozilla.org/en/privacynotincluded/}} for popular consumer devices and apps to highlight problematic privacy and security practices.
\newline

\noindent\textbf{Security of SSO login options.}
Across different participants in our study, both security and privacy of personal information were significant factors for choosing and avoiding both SSO and non-SSO login options.
Beyond our study, SSO protocols such as OAuth simplify security for RPs by outsourcing password management, but also complicate security because implementing OAuth 2.0 securely is challenging~\cite{parecki2019time}, e.g., requiring developers to understand various OAuth 2.0 extended features (in separate RFCs for a multitude of use cases and security mechanisms) and to decide which ones match specific web services.
Other research (e.g.,~\cite{ghasemisharif2018single,rahat2022cerberus,sun2012devil,zhou2014ssoscan,jannett2022distinct}) has found many vulnerabilities in RP implementations, raising security and privacy concerns about personal information in RP and IdP user accounts.

We encourage future work that explores the effects of security-related information on SSO and non-SSO login choices and how to help users make better-informed decisions with respect to (not only privacy-related factors as in our study but also) use of sites with security weaknesses, to complement our work herein focused on how displaying privacy-related information affects users' login decisions.

Besides improving the privacy (as studied herein) of OAuth systems, new browser extensions could explore the possibility of identifying known vulnerabilities in RP and IdP implementations to warn users before choosing SSO login options.
\newline

\noindent
Web SSO login systems have expanded from authentication systems to data-release authorization systems whereby users end up authorizing IdPs to release user personal information to RPs.
The volume of this personal information released to RPs increases with increased use of web SSO (and as users generate more such information over time); for this reason, we believe it has become more important to understand how factors studied herein including user perceptions, and the information displayed to users, influence their web SSO login decisions.
Our work is an early step in this direction.

\begin{acks}
We thank the anonymous referees for their helpful comments.
Morkonda acknowledges a scholarship from OGS.
Chiasson acknowledges NSERC for funding of an Arthur B. McDonald Fellowship and a Discovery Grant.  
Van Oorschot is Canada Research Chair in Authentication and Computer Security, and acknowledges funding from NSERC for the chair and a Discovery Grant.
\end{acks}

\bibliographystyle{abbrv}
\bibliography{references}

\begin{thebibliography}{10}

\bibitem{alaca2020comparative}
F.~Alaca and P.~C. van Oorschot.
\newblock {Comparative Analysis and Framework Evaluating Web Single Sign-on Systems}.
\newblock {\em {ACM Computing Surveys}}, 53(5):112:1--112:34, 2020.

\bibitem{balash2022security}
D.~G. Balash, X.~Wu, M.~Grant, I.~Reyes, and A.~J. Aviv.
\newblock {Security and Privacy Perceptions of Third-Party Application Access for Google Accounts}.
\newblock In {\em USENIX Security}, 2022.

\bibitem{bauer2013comparison}
L.~Bauer, C.~Bravo-Lillo, E.~Fragkaki, and W.~Melicher.
\newblock {A Comparison of Users' Perceptions of and Willingness to Use Google, Facebook, and Google+ Single-Sign-On Functionality}.
\newblock In {\em ACM Workshop on Digital Identity Management}, pages 25--36, 2013.

\bibitem{bonneau2012quest}
J.~Bonneau, C.~Herley, P.~C. van Oorschot, and F.~Stajano.
\newblock {The Quest to Replace Passwords: A Framework for Comparative Evaluation of Web Authentication Schemes}.
\newblock In {\em IEEE Symp. Security and Privacy}, 2012.

\bibitem{braun2012thematic}
V.~Braun and V.~Clarke.
\newblock {Thematic Analysis}.
\newblock In {\em APA handbook of research methods in psychology, Volume 2, Chapter 4}, pages 57--71. American Psychological Association, 2012.

\bibitem{cao2021large}
W.~Cao, C.~Xia, S.~T. Peddinti, D.~Lie, N.~Taft, and L.~M. Austin.
\newblock {A Large Scale Study of User Behavior, Expectations and Engagement with Android Permissions}.
\newblock In {\em USENIX Security}, pages 803--820, 2021.

\bibitem{chiasson2009multiple}
S.~Chiasson, A.~Forget, E.~Stobert, P.~C. van Oorschot, and R.~Biddle.
\newblock {Multiple Password Interference in Text Passwords and Click-Based Graphical Passwords}.
\newblock In {\em ACM CCS}, 2009.

\bibitem{google2022ExpandingADA}
B.~Davis.
\newblock {Expanding the App Defense Alliance}.
\newblock \url{https://security.googleblog.com/2022/12/app-defense-alliance-expansion.html}, December 15, 2022.

\bibitem{egelman2013my}
S.~Egelman.
\newblock {My Profile is My Password, Verify Me! The Privacy/Convenience Tradeoff of Facebook Connect}.
\newblock In {\em CHI}, page 2369–2378, 2013.

\bibitem{englehardt2016online}
S.~Englehardt and A.~Narayanan.
\newblock {Online Tracking: A 1-million-site Measurement and Analysis}.
\newblock In {\em ACM CCS}, 2016.

\bibitem{FacebookLogin}
Facebook.
\newblock {Facebook Login}.
\newblock \url{https://developers.facebook.com/docs/facebook-login/guides/permissions}, Accessed: April 7, 2023.

\bibitem{Facebook2023AppReview}
{Facebook Login}.
\newblock {Introduction - App Review}.
\newblock \url{https://developers.facebook.com/docs/app-review/introduction}, Accessed: April 25, 2023.

\bibitem{farooqi2020canarytrap}
S.~Farooqi, M.~Musa, Z.~Shafiq, and F.~Zaffar.
\newblock {Canarytrap: Detecting Data Misuse by Third-Party Apps on Online Social Networks}.
\newblock {\em Proceedings on Privacy Enhancing Technologies}, 2020(4):336--354, 2020.

\bibitem{felt2008privacy}
A.~Felt and D.~Evans.
\newblock {Privacy Protection for Social Networking APIs}.
\newblock {\em Web 2.0 Security and Privacy (W2SP)}, 2008.

\bibitem{felt2012how}
A.~P. Felt, S.~Egelman, M.~Finifter, D.~Akhawe, and D.~Wagner.
\newblock {How to Ask for Permission}.
\newblock In {\em Proceedings of the 7th USENIX Conference on Hot Topics in Security}, HotSec'12, page~7, 2012.

\bibitem{ghasemisharif2018single}
M.~Ghasemisharif, A.~Ramesh, S.~Checkoway, C.~Kanich, and J.~Polakis.
\newblock {O Single Sign-Off, Where Art Thou? An Empirical Analysis of Single Sign-On Account Hijacking and Session Management on the Web}.
\newblock In {\em USENIX Security}, 2018.

\bibitem{google2023OAuthAppVerification}
Google.
\newblock {OAuth API verification FAQs}.
\newblock \url{https://support.google.com/cloud/answer/9110914}, Accessed: April 25, 2023.

\bibitem{oauth2rfc}
D.~Hardt.
\newblock {RFC 6749: The OAuth 2.0 Authorization Framework}.
\newblock \url{https://tools.ietf.org/html/rfc6749}, 2012.

\bibitem{jannett2022distinct}
L.~Jannett, V.~Mladenov, C.~Mainka, and J.~Schwenk.
\newblock {DISTINCT: Identity Theft using In-Browser Communications in Dual-Window Single Sign-On}.
\newblock In {\em ACM CCS}, 2022.

\bibitem{jarpehult2022longitudinal}
O.~Järpehult, F.~J. Ågren, M.~Bäckström, L.~Hallonqvist, and N.~Carlsson.
\newblock {A Longitudinal Characterization of the Third-Party Authentication Landscape}.
\newblock In {\em International Federation for Information Processing (IFIP) Networking}, 2022.

\bibitem{liu2016follow}
B.~Liu, M.~S. Andersen, F.~Schaub, H.~Almuhimedi, S.~A. Zhang, N.~Sadeh, Y.~Agarwal, and A.~Acquisti.
\newblock {Follow My Recommendations: A Personalized Privacy Assistant for Mobile App Permissions}.
\newblock In {\em SOUPS}, pages 27--41, June 2016.

\bibitem{morkonda2021empirical}
S.~G. Morkonda, S.~Chiasson, and P.~C. van Oorschot.
\newblock {Empirical Analysis and Privacy Implications in OAuth-based Single Sign-On Systems}.
\newblock In {\em Workshop on Privacy in the Electronic Society}, 2021.

\bibitem{morkonda2022ssoprivateeye}
S.~G. Morkonda, S.~Chiasson, and P.~C. van Oorschot.
\newblock {SSOPrivateEye: Timely Disclosure of Single Sign-On Privacy Design Differences}, 2022.
\newblock Manuscript. A preliminary version is at: \url{https://arxiv.org/abs/2209.04490}.

\bibitem{parecki2019time}
A.~Parecki.
\newblock {It's Time for OAuth 2.1}.
\newblock \url{https://aaronparecki.com/2019/12/12/21/its-time-for-oauth-2-dot-1}, 2019.

\bibitem{polites2012shackled}
G.~L. Polites and E.~Karahanna.
\newblock {Shackled to the Status Quo: The Inhibiting Effects of Incumbent System Habit, Switching Costs, and Inertia on New System Acceptance}.
\newblock {\em MIS quarterly}, pages 21--42, 2012.

\bibitem{rahat2022cerberus}
T.~A. Rahat, Y.~Feng, and Y.~Tian.
\newblock {Cerberus: Query-driven Scalable Security Checking for OAuth Service Provider Implementations}.
\newblock In {\em ACM CCS}, 2022.

\bibitem{robinson2014cognitive}
N.~Robinson and J.~Bonneau.
\newblock {Cognitive Disconnect: Understanding Facebook Connect Login Permissions}.
\newblock In {\em ACM COSN}, 2014.

\bibitem{fb2018ssobreach}
G.~Rosen.
\newblock {Facebook Security Update - Security issue affecting almost 50 million accounts}.
\newblock \url{https://about.fb.com/news/2018/09/security-update/}, September 28, 2018.

\bibitem{openIdConnect1Spec}
N.~Sakimura, J.~Bradley, M.~B. Jones, B.~de~Medeiros, and C.~Mortimore.
\newblock {OpenID Connect Core 1.0}.
\newblock \url{https://openid.net/specs/openid-connect-core-1\_0.html}, 2014.

\bibitem{Apple2023PrivateEmailRelay}
{Sign in with Apple}.
\newblock {Communicating using the Private Email Relay Service}.
\newblock \url{https://developer.apple.com/documentation/sign_in_with_apple/sign_in_with_apple_js/communicating_using_the_private_email_relay_service}, Accessed: April 25, 2023.

\bibitem{SignInWithGoogle}
{Sign in with Google}.
\newblock {Setup - Configure your OAuth Consent Screen}.
\newblock \url{https://developers.google.com/identity/gsi/web/guides/get-google-api-clientid}, Accessed: April 7. 2023.

\bibitem{sun2012devil}
S.-T. Sun and K.~Beznosov.
\newblock {The Devil is in the (Implementation) Details: An Empirical Analysis of OAuth SSO Systems}.
\newblock In {\em ACM CCS}, 2012.

\bibitem{sun2011makes}
S.-T. Sun, E.~Pospisil, I.~Muslukhov, N.~Dindar, K.~Hawkey, and K.~Beznosov.
\newblock {What Makes Users Refuse Web Single Sign-On? An Empirical Investigation of OpenID}.
\newblock In {\em SOUPS}, 2011.

\bibitem{ur2015added}
B.~Ur, F.~Noma, J.~Bees, S.~M. Segreti, R.~Shay, L.~Bauer, N.~Christin, and L.~F. Cranor.
\newblock {"I Added`!'at the End to Make It Secure": Observing Password Creation in the Lab}.
\newblock In {\em SOUPS}, 2015.

\bibitem{Na2011Third}
N.~Wang, H.~Xu, and J.~Grossklags.
\newblock Third-party apps on facebook: Privacy and the illusion of control.
\newblock In {\em ACM Symposium on Computer Human Interaction for Management of Information Technology}, 2011.

\bibitem{weinshel2019oh}
B.~Weinshel, M.~Wei, M.~Mondal, E.~Choi, S.~Shan, C.~Dolin, M.~L. Mazurek, and B.~Ur.
\newblock {Oh, the Places You've Been! User Reactions to Longitudinal Transparency About Third-Party Web Tracking and Inferencing}.
\newblock In {\em ACM CCS}, pages 149--166, 2019.

\bibitem{wijesekera2017feasibility}
P.~Wijesekera, A.~Baokar, L.~Tsai, J.~Reardon, S.~Egelman, D.~Wagner, and K.~Beznosov.
\newblock {The Feasibility of Dynamically Granted Permissions: Aligning Mobile Privacy with User Preferences}.
\newblock In {\em IEEE Symp. Security and Privacy}, pages 1077--1093, 2017.

\bibitem{zhou2014ssoscan}
Y.~Zhou and D.~Evans.
\newblock {SSOScan: Automated Testing of Web Applications for Single Sign-On Vulnerabilities}.
\newblock In {\em USENIX Security}, 2014.

\end{thebibliography}

\appendix

\section{Study Questionnaire}
\label{sec.app.survey}
\subsection*{Basic demographic and SSO usage questions}
\begin{enumerate}[left=0pt .. \parindent,label=\arabic{enumi}.]
\item What is your Prolific ID? \_\_\_\_\_\_
\newline

\item What is your gender?
    \begin{itemize}[left=0.5em]
        \item Man
        \item Woman
        \item Non-binary
        \item Not listed above [please specify]: \_\_\_\_\_\_
        \item Prefer not to answer
   \end{itemize}
\medskip

\item Please enter your age in years. If you prefer not to answer, enter 100 \_\_\_\_\_\_
\newline
\item What is your highest level of education (completed or currently enrolled)?
    \begin{itemize}[left=0.5em]
        \item High school diploma or lower
        \item College or associate degree
        \item Bachelor's degree
        \item Advanced degree (e.g., Master's, PhD)
        \item Prefer not to answer
        \item Other [please specify]: \_\_\_\_\_\_
    \end{itemize}
\medskip    

\item How many different websites do you log into on a typical day? [0-2], [3-5], [6-8], [9-11], [12+], [prefer not to answer]
\newline
\item Single Sign-On (SSO) includes services that allow you to login to a website using another account (e.g., "Sign in with Facebook", "Login with Google", "Continue with Apple". etc.).
\medskip

Have you ever used single sign-on (SSO) to login to a website? [Yes], [No], [Unsure], [Prefer not to answer]
\bigskip

\begin{center}
\textit{<page break -- participants could not change earlier responses>}
\end{center}
\end{enumerate}

\begin{center}
\textit{<Participants were shown one of Group A, B, C, or D>}
\end{center}

\subsection*{Group A - Airbnb.ca}
\begin{enumerate}[left=0pt .. \parindent,label=\arabic{enumi}.]
    \item If you were to login to Airbnb.ca, which login option would you choose?
    
    \frame{\includegraphics[width=0.5\columnwidth]{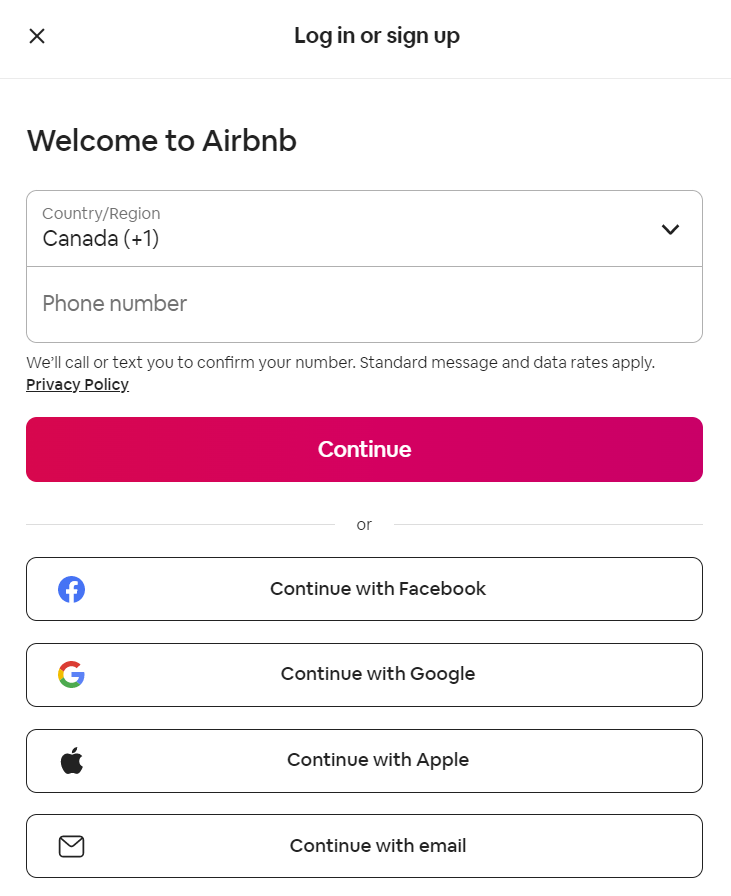}}
    \begin{itemize}[left=0.5em]
        \item Continue with Phone number
        \item Continue with Facebook
        \item Continue with Google
        \item Continue with Apple
        \item Continue with email
        \item Prefer not to answer
        \item I would take some other action [please specify]: \_\_\_\_\_\_
    \end{itemize}
    \bigskip

    \item Why did you choose this option? If you prefer not to answer, enter N/A (No Answer) \_\_\_\_\_\_
    \bigskip

    \begin{center}
    \textit{<page break -- participants could not change earlier responses>}
    \end{center}
    \medskip

    \item If the following information was available when logging in to Airbnb.ca, which login option would you choose?
    \begin{figure}[ht]
    \centering
        \subfloat{
            \frame{\includegraphics[width=0.34\columnwidth]{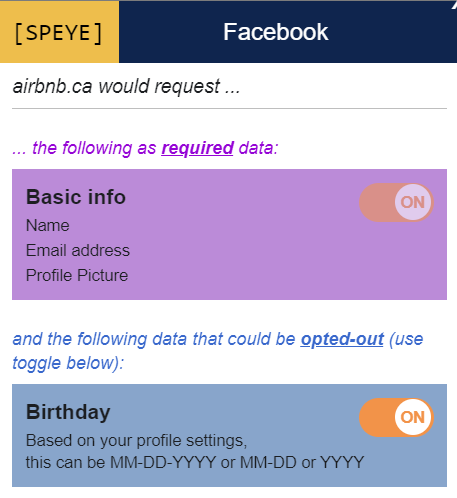}}}
        \subfloat{
            \frame{\includegraphics[width=0.34\columnwidth]{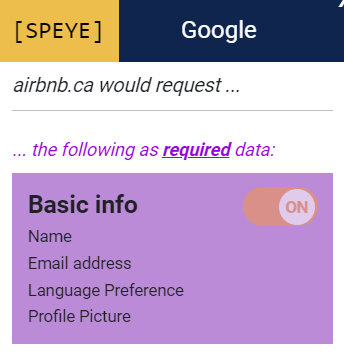}}}
        \subfloat{
            \frame{\includegraphics[width=0.34\columnwidth]{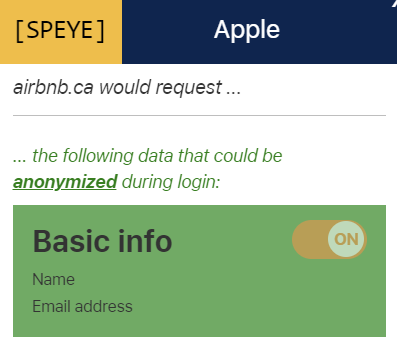}}}
    \end{figure}
    \begin{center}
    \textit{<Airbnb.ca login prompt from Q1 displayed again here>}
    \end{center}
    \medskip
    \begin{itemize}[left=0.5em]
        \item Continue with Phone number
        \item Continue with Facebook (with all requested data included)
        \item Continue with Facebook (with birthday data opted out)
        \item Continue with Google
        \item Continue with Apple (with real name and email address) 
        \item Continue with Apple (with anonymized name and email address)
        \item Continue with email
        \item Prefer not to answer
        \item I would take some other action [please specify]: \_\_\_\_\_\_
    \end{itemize}
    \bigskip

    \item Why did you choose this option? If you prefer not to answer, enter N/A (No Answer) \_\_\_\_\_\_
    \bigskip

    \begin{center}
    \textit{<page break -- participants could not change earlier responses>}
    \end{center}
    \bigskip

    \item Indicate how much each of the following impacted your chosen login option in the previous question.
    \smallskip\newline
    \textit{Reminder: You selected <copy Q3 response> in the previous question}
    \smallskip\newline
    [\textit{Displayed only if ``Prefer not to answer'' was not selected in Q3}]
    \smallskip\newline
    [\textit{Likert scale responses - 1 (strongly disagree) to 5 (strongly agree)}]
    \medskip
    
    \begin{enumerate}[left=0.5em,label=\alph{enumii}.,noitemsep]
        \item I chose this login option because I have an existing account with the SSO provider. $\dagger$ \smallskip
        \item I chose this login option because I trust the SSO provider. $\dagger$ \smallskip
        \item I chose this login option because it was listed first in the login prompt. \smallskip
        \item I chose this login option because it requested less data than other options. \smallskip
        \item I chose this login option because it lets me opt-out of requested data. $\dagger$ \smallskip
        \item I chose this login option because it lets me anonymize my data. $\dagger$ \smallskip
        \item I don’t want to use SSO on Airbnb.ca
    \end{enumerate}
    \medskip

    [$\dagger$ \textit{Displayed only if an SSO login option was selected in Q3}]
    \bigskip

    \item Do you have an existing account with Airbnb?
    \newline
    [Yes] [No] [I don't know] [Prefer not to answer]
    \medskip

    \begin{center}
    \textit{<page break -- participants could not change earlier responses>}
    \end{center}
\end{enumerate}

\subsection*{Group B - Rakuten.com}
\begin{enumerate}[left=0pt .. \parindent,label=\arabic{enumi}.]
    \item If you were to login to Rakuten.com, which login option would you choose?
    
    \frame{\includegraphics[width=0.45\columnwidth]{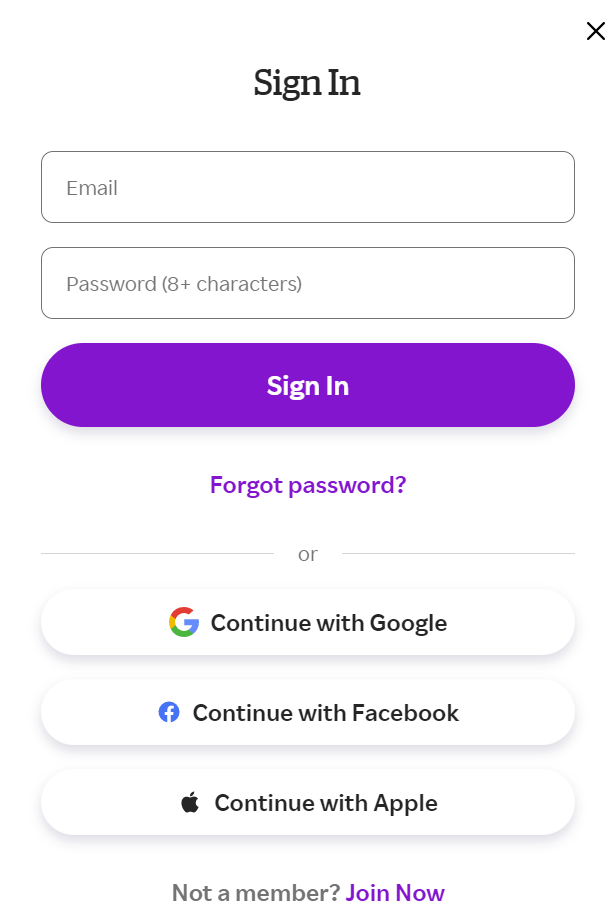}}
    \begin{itemize}[left=0.5em]
        \item Sign in using Email and Password
        \item Continue with Google
        \item Continue with Facebook
        \item Continue with Apple
        \item Prefer not to answer
        \item I would take some other action [please specify]: \_\_\_\_\_\_
    \end{itemize}
    \bigskip

    \item Why did you choose this option? If you prefer not to answer, enter N/A (No Answer) \_\_\_\_\_\_
    \bigskip

    \begin{center}
    \textit{<page break -- participants could not change earlier responses>}
    \end{center}
    \bigskip

    \item If the following information was available when logging in to Rakuten.com, which login option would you choose?
    \begin{figure}[ht]
    \centering
        \subfloat{
            \frame{\includegraphics[width=0.34\columnwidth]{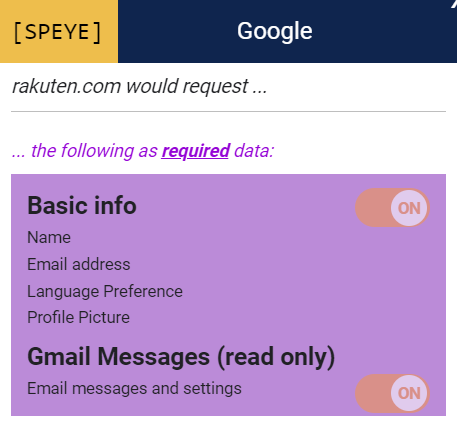}}}
        \subfloat{
            \frame{\includegraphics[width=0.34\columnwidth]{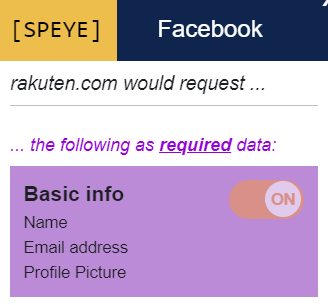}}}
        \subfloat{
            \frame{\includegraphics[width=0.34\columnwidth]{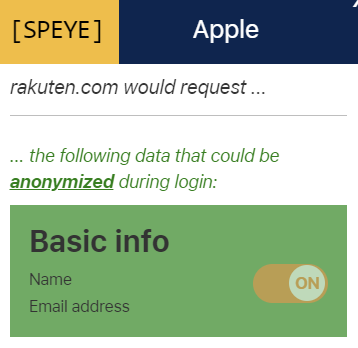}}}
    \end{figure}
    \begin{center}
    \textit{<Rakuten.com login prompt from Q1 displayed again here>}
    \end{center}
    \medskip
    \begin{itemize}[left=0.5em]
        \item Sign in using Email and Password
        \item Continue with Google
        \item Continue with Facebook
        \item Continue with Apple (with real name and email address)
        \item Continue with Apple (with anonymized name and email address)
        \item Prefer not to answer
        \item I would take some other action [please specify]: \_\_\_\_\_\_
    \end{itemize}
    \bigskip

    \item Why did you choose this option? If you prefer not to answer, enter N/A (No Answer) \_\_\_\_\_\_
    \bigskip

    \begin{center}
    \textit{<page break -- participants could not change earlier responses>}
    \end{center}
    \bigskip

    \item Indicate how much each of the following impacted your chosen login option in the previous question.
    \smallskip\newline
    \textit{Reminder: You selected <copy Q3 response> in the previous question}
    \smallskip\newline
    [\textit{Displayed only if ``Prefer not to answer'' was not selected in Q3}]
    \smallskip\newline
    [\textit{Likert scale responses - 1 (strongly disagree) to 5 (strongly agree)}]
    \medskip
    
    \begin{enumerate}[left=0.5em,label=\alph{enumii}.,noitemsep]
        \item I chose this login option because I have an existing account with the SSO provider. $\dagger$ \smallskip
        \item I chose this login option because I trust the SSO provider. $\dagger$ \smallskip
        \item I chose this login option because it was listed first in the login prompt. \smallskip
        \item I chose this login option because it requested less data than other options. \smallskip
        \item I chose this login option because it lets me opt-out of requested data. $\dagger$ \smallskip
        \item I chose this login option because it lets me anonymize my data. $\dagger$ \smallskip
        \item I don’t want to use SSO on Rakuten.com
    \end{enumerate}
    \medskip

    [$\dagger$ \textit{Displayed only if an SSO login option was selected in Q3}]
    \bigskip

    \item Do you have an existing account with Rakuten?
    \newline
    [Yes] [No] [I don't know] [Prefer not to answer]
    \bigskip

    \begin{center}
    \textit{<page break -- participants could not change earlier responses>}
    \end{center}
\end{enumerate}

\subsection*{Group C - NYTimes.com}
\begin{enumerate}[left=0pt .. \parindent,label=\arabic{enumi}.]
    \item If you were to login to NYTimes.com, which login option would you choose?
    
    \frame{\includegraphics[width=0.5\columnwidth]{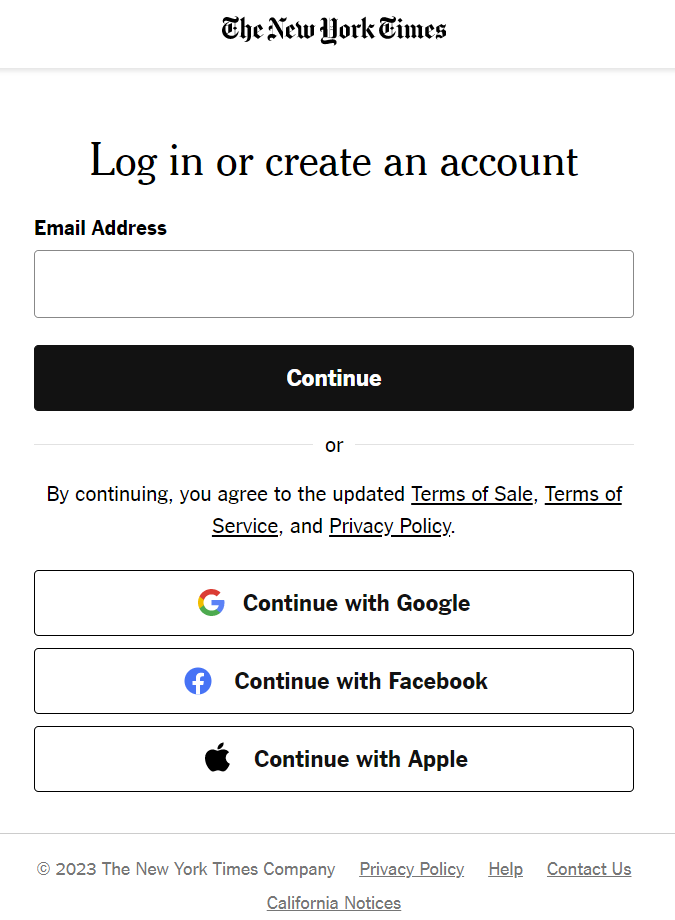}}
    \begin{itemize}[left=0.5em]
        \item Log in with Email Address
        \item Continue with Google
        \item Continue with Facebook
        \item Continue with Apple
        \item Prefer not to answer
        \item I would take some other action [please specify]: \_\_\_\_\_\_
    \end{itemize}
    \bigskip

    \item Why did you choose this option? If you prefer not to answer, enter N/A (No Answer) \_\_\_\_\_\_
    \bigskip

    \begin{center}
    \textit{<page break -- participants could not change earlier responses>}
    \end{center}
    \bigskip

    \item If the following information was available when logging in to NYTimes.com, which login option would you choose?
    \begin{figure}[ht]
    \centering
        \subfloat{
            \frame{\includegraphics[width=0.34\columnwidth]{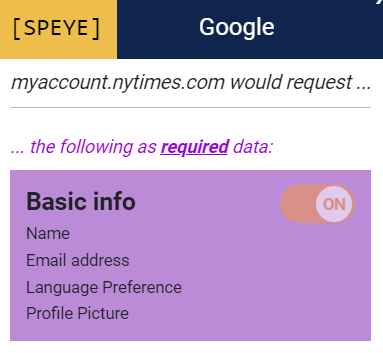}}}
        \subfloat{
            \frame{\includegraphics[width=0.34\columnwidth]{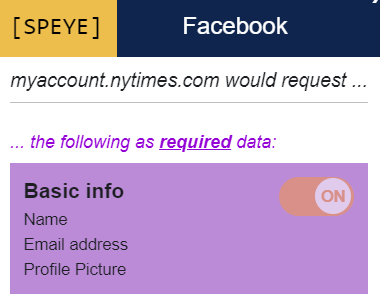}}}
        \subfloat{
            \frame{\includegraphics[width=0.34\columnwidth]{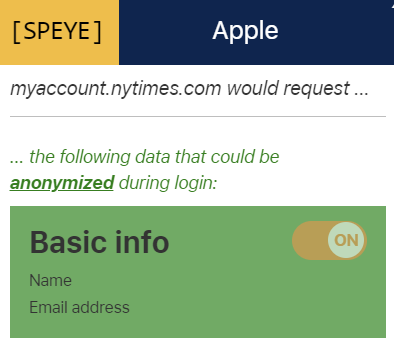}}}
    \end{figure}
    \begin{center}
    \textit{<NYTimes.com login prompt from Q1 displayed again here>}
    \end{center}
    \medskip
    \begin{itemize}[left=0.5em]
        \item Log in with Email Address
        \item Continue with Google
        \item Continue with Facebook
        \item Continue with Apple (with real name and email address)
        \item Continue with Apple (with anonymized name and email address)
        \item Prefer not to answer
        \item I would take some other action [please specify]: \_\_\_\_\_\_
    \end{itemize}
    \bigskip

    \item Why did you choose this option? If you prefer not to answer, enter N/A (No Answer) \_\_\_\_\_\_
    \bigskip

    \begin{center}
    \textit{<page break -- participants could not change earlier responses>}
    \end{center}
    \bigskip

    \item Indicate how much each of the following impacted your chosen login option in the previous question.
    \smallskip\newline
    \textit{Reminder: You selected <copy Q3 response> in the previous question}
    \smallskip\newline
    [\textit{Displayed only if ``Prefer not to answer'' was not selected in Q3}]
    \smallskip\newline
    [\textit{Likert scale responses - 1 (strongly disagree) to 5 (strongly agree)}]
    \medskip
    
    \begin{enumerate}[left=0.5em,label=\alph{enumii}.,noitemsep]
        \item I chose this login option because I have an existing account with the SSO provider. $\dagger$ \smallskip
        \item I chose this login option because I trust the SSO provider. $\dagger$ \smallskip
        \item I chose this login option because it was listed first in the login prompt. \smallskip
        \item I chose this login option because it requested less data than other options. \smallskip
        \item I chose this login option because it lets me opt-out of requested data. $\dagger$ \smallskip
        \item I chose this login option because it lets me anonymize my data. $\dagger$ \smallskip
        \item I don’t want to use SSO on NYTimes.com
    \end{enumerate}
    \medskip

    [$\dagger$ \textit{Displayed only if an SSO login option was selected in Q3}]
    \bigskip

    \item Do you have an existing account with NYTimes?
    \newline
    [Yes] [No] [I don't know] [Prefer not to answer]
    \bigskip

    \begin{center}
    \textit{<page break -- participants could not change earlier responses>}
    \end{center}
\end{enumerate}

\subsection*{Group D - CBC.ca}
\begin{enumerate}[left=0pt .. \parindent,label=\arabic{enumi}.]
    \item If you were to login to CBC.ca, which login option would you choose?
    
    \frame{\includegraphics[width=0.5\columnwidth]{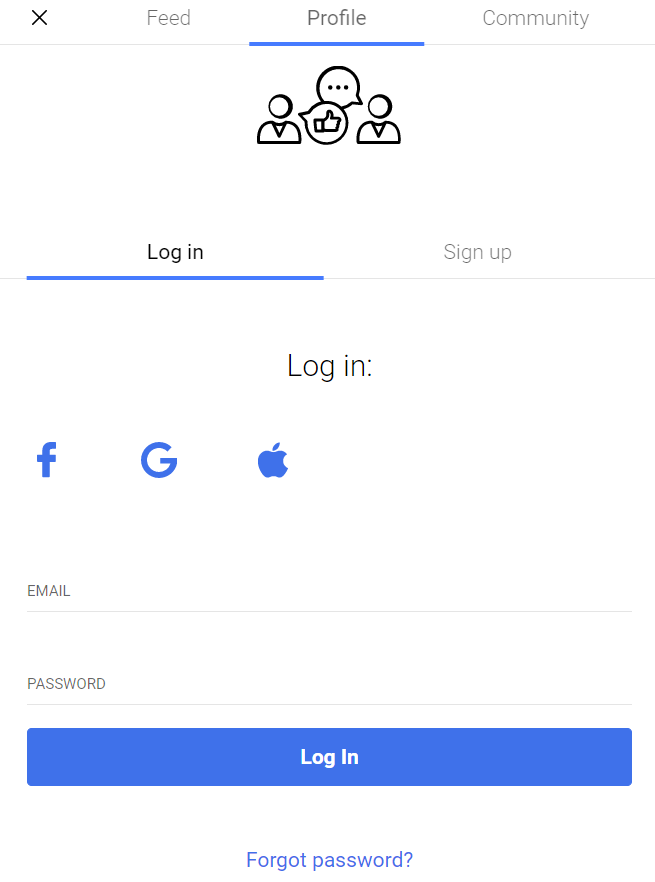}}
    \begin{itemize}[left=0.5em]
        \item Log in with Facebook
        \item Log in with Google
        \item Log in with Apple
        \item Log in with Email and Password
        \item Prefer not to answer
        \item I would take some other action [please specify]: \_\_\_\_\_\_
    \end{itemize}
    \bigskip

    \item Why did you choose this option? If you prefer not to answer, enter N/A (No Answer) \_\_\_\_\_\_
    \bigskip

    \begin{center}
    \textit{<page break -- participants could not change earlier responses>}
    \end{center}
    \bigskip

    \item If the following information was available when logging in to CBC.ca, which login option would you choose?
    \begin{figure}[ht]
    \centering
        \subfloat{
            \frame{\includegraphics[width=0.34\columnwidth]{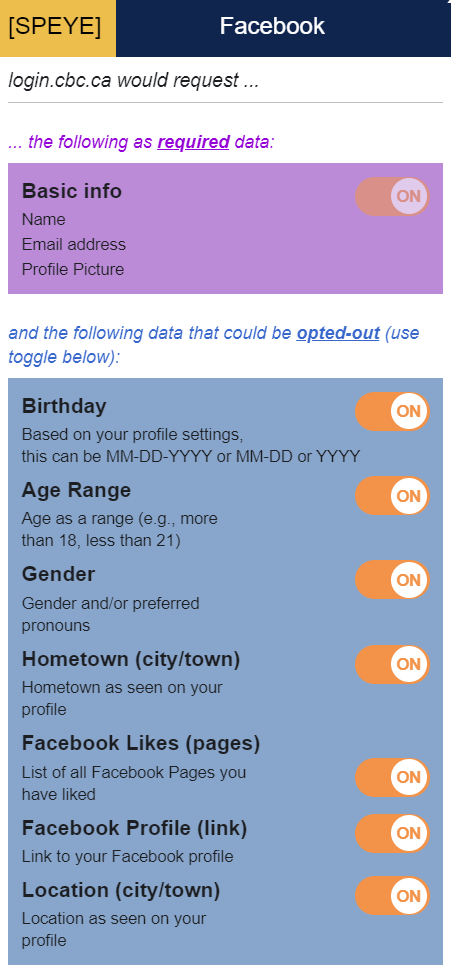}}}
        \subfloat{
            \frame{\includegraphics[width=0.34\columnwidth]{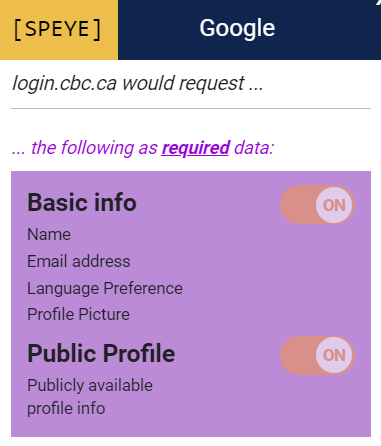}}}
        \subfloat{
            \frame{\includegraphics[width=0.34\columnwidth]{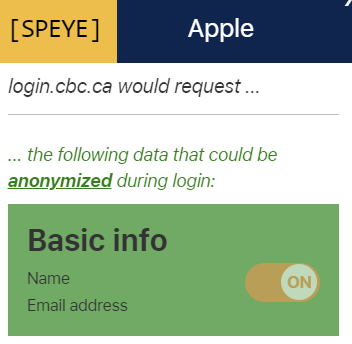}}}
    \end{figure}
    \begin{center}
    \textit{<CBC.ca login prompt from Q1 displayed again here>}
    \end{center}
    \medskip
    \begin{itemize}[left=0.5em]
        \item Log in with Facebook (with all requested data included)
        \item Log in with Facebook (with 1+ permissions opted-out)
        \item Log in with Google
        \item Log in with Apple (with real name and email address)
        \item Log in with Apple (with anonymized name and email address)
        \item Log in with Email and Password
        \item Prefer not to answer
        \item I would take some other action [please specify]: \_\_\_\_\_\_
    \end{itemize}
    \bigskip

    \item Why did you choose this option? If you prefer not to answer, enter N/A (No Answer) \_\_\_\_\_\_
    \bigskip

    \begin{center}
    \textit{<page break -- participants could not change earlier responses>}
    \end{center}
    \bigskip

    \item Indicate how much each of the following impacted your chosen login option in the previous question.
    \smallskip\newline
    \textit{Reminder: You selected <copy Q3 response> in the previous question}
    \smallskip\newline
    [\textit{Displayed only if ``Prefer not to answer'' was not selected in Q3}]
    \smallskip\newline
    [\textit{Likert scale responses - 1 (strongly disagree) to 5 (strongly agree)}]
    \medskip
    
    \begin{enumerate}[left=0.5em,label=\alph{enumii}.,noitemsep]
        \item I chose this login option because I have an existing account with the SSO provider. $\dagger$ \smallskip
        \item I chose this login option because I trust the SSO provider. $\dagger$ \smallskip
        \item I chose this login option because it was listed first in the login prompt. \smallskip
        \item I chose this login option because it requested less data than other options. \smallskip
        \item I chose this login option because it lets me opt-out of requested data. $\dagger$ \smallskip
        \item I chose this login option because it lets me anonymize my data. $\dagger$ \smallskip
        \item I don’t want to use SSO on CBC.ca
    \end{enumerate}
    \medskip

    [$\dagger$ \textit{Displayed only if an SSO login option was selected in Q3}]
    \bigskip

    \item Do you have an existing account with CBC?
    \newline
    [Yes] [No] [I don't know] [Prefer not to answer]
    \bigskip

    \begin{center}
    \textit{<page break -- participants could not change earlier responses>}
    \end{center}
\end{enumerate}

\subsection*{Single Sign-On Use}
\begin{enumerate}[left=0pt .. \parindent,label=\arabic{enumi}.]

\item Outside of work, on how many different websites do you use single sign-on (SSO) login?
\begin{multicols}{2}
\begin{itemize}[left=0.5em]
    \item I don't use SSO
    \item 1-5 websites
    \item 6-10 websites
    \item 11-15 websites
    \item 16-20 websites
    \item 21+ websites
    \item Prefer not to answer
\end{itemize}
\end{multicols}
\bigskip

\item Answer the following questions about using single sign-on (SSO) login outside of work. \medskip

[\textit{Likert scale responses - 1 (strongly disagree) to 5 (strongly agree)}]
\medskip

\begin{enumerate}[left=0.5em,label=\alph{enumii}.]
    \item SSO is my preferred login method
    \item I find it easy to use SSO login
    \item Using SSO login helps reduce the number of passwords I need to manage.
    \item I pay close attention to the types of data requested during SSO login.
    \item The ability to opt-out of data requested during SSO login is important to me.
\end{enumerate}
\bigskip

\item Outside of work, with which of following do you have an account? [Select all that apply]

\begin{multicols}{2}
\begin{multiselect}[left=0.5em]
    \item Apple
    \item Facebook
    \item GitHub
    \item Google
    \item LinkedIn
    \item Microsoft 
    \item Twitter
    \item Vk 
    \item Yahoo!
    \item Prefer not to answer
    \item None of these
\end{multiselect}
\end{multicols}
\bigskip

\item Other than for SSO login, how often do you use applications and/or services related to accounts from these providers (selected in the previous question) outside of work?
\medskip

[\textit{Displayed if ``Prefer not to answer'' or ``None of these'' was not selected in Q3}]

[\textit{Likert scale rows for each choice selected by the participant in Q3}] \medskip

\begin{tabular}{c@{\hspace{6pt}}c@{\hspace{6pt}}c@{\hspace{6pt}}c@{\hspace{6pt}}c@{\hspace{6pt}}}
     At least once & At least once & At least once & At least once &  \\
    a day & a week & a month & a year & Never \\
     $\circ$ & $\circ$ & $\circ$ & $\circ$ & $\circ$ \\
\end{tabular}
\bigskip

\item Which of the following accounts have you used at least once for SSO login outside of work?

\begin{multicols}{2}
\begin{multiselect}[left=0.5em]
    \item I don't use SSO
    \item Apple
    \item Facebook
    \item GitHub 
    \item Google 
    \item LinkedIn
    \item Microsoft
    \item Twitter
    \item Vk
    \item Yahoo!
    \item Prefer not to answer 
    \item Other [please specify]: \_\_\_\_\_\_
\end{multiselect}
\end{multicols}

\end{enumerate}

\end{document}